\documentclass[12pt]{article}
\usepackage{epsfig}
\usepackage{psfrag}
\usepackage{latexsym}
\usepackage{indentfirst}
\usepackage{fancyhdr}
\usepackage{amssymb}
\usepackage{amsmath}
\usepackage{amsfonts}
\usepackage{cite}
\usepackage{bbold}
\usepackage{color}
\usepackage[footnotesize]{caption2}
\usepackage{graphicx}
\usepackage[center,footnotesize,hang]{subfigure}
\usepackage{url}
\usepackage{amscd}

\makeatletter

\newcommand{\Rmnum}[1]{\expandafter\@slowromancap\romannumeral #1@}
\makeatother

\newcommand{\vphi}{\varphi}

\newcommand{\pa}{\partial}

\newcommand{\td}{\tilde}

\newcommand{\la}{\label}
\newcommand{\bm}[1]{\mbox{\boldmath{$#1$}}}
\renewcommand{\theequation}{\thesection.\arabic{equation}}

\def\be{\begin{equation}}
\def\ee{\end{equation}}
\def\bc{\begin{center}}
\def\ec{\end{center}}
\def\bea{\begin{eqnarray}}
\def\eea{\end{eqnarray}}
\def\la{\label}

\def\nn{\nonumber}

\def\be{\begin{equation}}
\def\ee{\end{equation}}
\def\bc{\begin{center}}
\def\ec{\end{center}}
\def\bea{\begin{eqnarray}}
\def\eea{\end{eqnarray}}

\def\nn{\nonumber}

\catcode`@=11
\def\marginnote#1{}
\newcount\hour
\newcount\minute
\newtoks\amorpm
\hour=\time\divide\hour by60
\minute=\time{\multiply\hour by60 \global\advance\minute by-\hour}
\edef\standardtime{{\ifnum\hour<12 \global\amorpm={am}%
        \else\global\amorpm={pm}\advance\hour by-12 \fi
        \ifnum\hour=0 \hour=12 \fi
        \number\hour:\ifnum\minute<10 0\fi\number\minute\the\amorpm}}
\edef\militarytime{\number\hour:\ifnum\minute<10 0\fi\number\minute}
\def\draftlabel#1{{\@bsphack\if@filesw {\let\thepage\relax
   \xdef\@gtempa{\write\@auxout{\string
      \newlabel{#1}{{\@currentlabel}{\thepage}}}}}\@gtempa
   \if@nobreak \ifvmode\nobreak\fi\fi\fi\@esphack}
        \gdef\@eqnlabel{#1}}
\def\@eqnlabel{}
\def\@vacuum{}
\def\draftmarginnote#1{\marginpar{\raggedright\scriptsize\tt#1}}
\def\draft{\oddsidemargin 0.0truein
        \def\@oddfoot{\sl preliminary draft \hfil
        \rm\thepage\hfil\sl\today\quad\militarytime}
        \let\@evenfoot\@oddfoot \overfullrule 3pt
        \let\label=\draftlabel
        \let\marginnote=\draftmarginnote
   \def\@eqnnum{(\theequation)\rlap{\kern\marginparsep\tt\@eqnlabel}%
\global\let\@eqnlabel\@vacuum}  }
\catcode`@=12
\begin{document}
\begin{titlepage}
\vspace*{-1cm}
\phantom{hep-ph/***}

\hfill{USTC-ICTS-14-15}

\vskip 2.5cm
\begin{center}
{\Large\bf  Implication of Spatial and Temporal Variations of the Fine-Structure Constant    }
\end{center}
\vskip 0.2  cm
\vskip 0.5  cm
\begin{center}
{\large Sze-Shiang Feng \footnote{Email: de.sitter.universe@gmail.com},
Mu-Lin Yan}~\footnote{Email: mlyan@ustc.edu.cn; Corresponding author.}
\\
\vskip .3cm {\it Interdisciplinary Center for Theoretical Study,}
\\
 {\it Department of Modern Physics,}
\\
{\it University of Science and Technology of China, Hefei, Anhui
230026, China}
\end{center}
\vskip 0.7cm
\begin{abstract}
\noindent

\noindent Temporal and spatial variation of fine-structure constant $\alpha\equiv e^2/\hbar c$ in cosmology has been reported in analysis of combination Keck and VLT data. This paper studies this variation based on consideration of basic spacetime symmetry in physics. Both laboratory $\alpha_0$ and distant $\alpha_z$ are deduced from relativistic spectrum equations of atoms (e.g.,hydrogen atom) defined in inertial reference system. When Einstein's $\Lambda\neq 0$, the metric of local inertial reference systems in SM of cosmology is Beltrami metric instead of Minkowski, and the basic spacetime symmetry has to be de Sitter (dS) group. The corresponding special relativity (SR) is dS-SR. A model based on dS-SR is suggested. Comparing the predictions on $\alpha$-varying with the data, the parameters are determined. The best-fit dipole mode in $\alpha$'s spatial varying is reproduced by this dS-SR model. $\alpha$-varyings in whole sky is also studied. The results are generally in agreement with the estimations of observations. The main conclusion is that the phenomenon of $\alpha$-varying cosmologically with dipole mode dominating is due to the de Sitter (or anti de Sitter) spacetime symmetry with a Minkowski point in an extended special relativity called de Sitter invariant special relativity (dS-SR)
developed by Dirac-In\"{o}n\"{u}-Wigner-G\"{u}rsey-Lee-Lu-Zou-Guo.

\end{abstract}
\vskip0.2in

\noindent PACS numbers: 06.20.Jr, 95.30.Sf, 03.65.Pm, 98.62.Ra, 95.36.+x\\
Key words: Fine-structure constant varying; Spacetime symmetry in Special Relativity; Dirac equation of Hydrogen atom; Friedmann-Robertson-Walker (FRW) Universe; Local inertial coordinate systems.

\end{titlepage}
\setcounter{footnote}{1}
 \vskip2truecm
\section{Introduction}
Temporal and spatial variation of fundamental constants is a possibility,
or even a necessity, in an expanding Universe (see, \cite{dirac37}, and a review of \cite{Uzan}).
A change in the fine
structure constant  $\alpha\equiv e^2/(\hbar c)$ could be detected via shifts in the frequencies of
atomic transitions in quasar absorption systems.  Recent analysis of a combined sample of quasar absorption
line spectra obtained using UVES (the Ultraviolet
and Visual Echelle Spectrograph) on VLT (the Very Large Telescope
) and HIRES (the High Resolution Echelle
Spectrometer) on the Keck Telescope have provided hints
of a spatial variation of the fine structure constant $(\alpha)$,
which is well represented by an angular dipole model \cite{Webb1} \cite{Webb0}. That is, $\alpha$ could be
smaller  in one direction in the sky yet larger in the opposite direction at the time of absorption.
Prior to that,  measurements of
possible time variations of the fine-structure constant was achieved by the same method \cite{Webb99, Webb992, Webb01, Webb03, Webb04} in the Keck telescope. It has been shown that the time variation of $\alpha$ exists. Thus, $\alpha$ is a ``constant" varying with both red-shift $z$ (or cosmologic time) and direction in the sky equatorial coordinates. Namely, $\alpha=\alpha_z(\mathbf{\Omega})$ where $\Omega$ indicates the direction in the sky.
Those direct measurements of
possible space-time variations of the fine-structure constant
are of utmost importance for a complete understanding
of fundamental physics.

A straightforward conjecture for this phenomenon is that the space-time function of $\alpha(x)$ may be thought as a scalar field $\vphi(x)$ or a function of $\vphi(x)$ in the spacetime with some suitable dynamics (see, e.g, \cite{Begen} \cite{SBM} \cite{Barrow}). The $\vphi(x)$ is a matter field and fills the Universe everywhere. Sometime one could call it dilaton-like scalar field. Along this way of thinking, authors of reference \cite{Olive} argued that the spatial variation of the fine structure constant $\alpha$ may be attributable to the domain wall of $\vphi(x)$ in the Universe.

In this present paper, we would like to present a matter-field-free scheme to answer the challenging questions such as why the fine-structure ``constant" $\alpha$ varies over space-time, and why spatial variation of $\alpha$
is well represented by an angular dipole mode. The scheme is still in the framework of standard cosmology and of the Special Relativity (SR) theory except that SR's spacetime symmetry will be extended. Concretely, we shall apply the de Sitter invariant Dirac equation to the distant hydrogen atom to explain such variations of $\alpha$ in cosmology. The calculations are based on the theory of the de Sitter invariant special relativity (dS-SR) developed by Dirac-In\"{o}n\"{u}-Wigner-G\"{u}rsey-Lee-Lu-Zou-Guo \cite{Dirac35,Wigner,Lee,look,Lu74,Guo1,Ours,Yan0,Yan01}. To study atom physics in dS-SR were firstly called for by P.A.M. Dirac in 1935 \cite{Dirac35}.

To show the fine-structure constant $\alpha$ is unvarying over space-time in the Standard Model (SM) of physics including cosmology, we examine the relativistic wave equation of an electron in hydrogen in SM of physics. First, we consider the laboratory atom. From the viewpoint of cosmology, the energy level $E$ of a free hydrogen atom in laboratory is determined by the Dirac equation in a local inertial coordinates system located at the Earth in the Universe described by Friedmann-Robertson-Walker (FRW) metric. The spacetime metric of the local inertial system is Minkowski metric:
\bea\la{1-01}
\{\eta_{\mu\nu}\}=\left( \begin{array}{lccr}
                           1&0&0&0\\
                           0&-1&0&0\\
                           0&0&-1&0\\
                           0&0&0&-1
                           \end{array}\right),
\eea
which is spacetime independent. $E$ satisfies Dirac spectrum equation:
\bea\la{1-02}
E\psi=\left(-i\hbar c \bm{\alpha}\cdot \nabla -{e^2\over r}+m_ec^2\beta\right)\psi,
\eea
where $\{\bm{\alpha}\equiv \alpha^1\mathbf{i}+\alpha^2\mathbf{j}+\alpha^3\mathbf{k},\;\beta\} $ are Dirac matrices, $\nabla\equiv (\pa/\pa x^1)\mathbf{i}+(\pa/\pa x^2)\mathbf{j}+(\pa/\pa x^3)\mathbf{k}$ and $r=\sqrt{(x^1)^2+(x^2)^2+(x^3)^2}$. This matrix-differential equation is integrable and the solution of the eigenvalue $E$ is (see, e.g., \cite{Strange})
\begin{eqnarray}\label{solution1}
E\equiv W_{n,\kappa}&=&m_e c^2\left(1+{\alpha^2 \over
(n-|\kappa |+s)^2} \right)^{-1/2} \\ \nonumber
&&\alpha\equiv {e^2\over \hbar c},~~~~|\kappa|=(j+1/2)=1,\;2,\;3\;\cdots \\
\nonumber &&s=\sqrt{\kappa^2-\alpha^2},~~~~n=1,\;2,\;3\;\cdots.
\end{eqnarray}
We keep in mind that the coefficients of operators $-i\bm{\alpha}\cdot\nabla$ and $-1/r$ in Eq.(\ref{1-02}) are $\hbar c$ and $e^2$ respectively, and their ratio is the definition of $\alpha$ (see Eq(\ref{solution1})).

Next, we consider a distant atom of hydrogen located on the light-cone of FRW-Universe (see Fig.\ref{Fig1}), i.e., the nucleus coordinate is $Q^\mu(z)\equiv \{Q^0,\;\mathbf{Q}\}$, and electron's is $L^\mu(z)\equiv \{L^0,\;\mathbf{L}\}$. Noting that the metric of the local inertial coordinate system at $Q^\mu$ in FRW-Universe is still $\eta_{\mu\nu}$ (\ref{1-01}) because of the spacetime-independency of $\eta_{\mu\nu}$ and denoting $L^\mu(z)-Q^\mu(z)\equiv x'^\mu$, the electron wave equation in the distant atom reads
\bea\la{1-04}
E\psi=\left(-i\hbar c \bm{\alpha}\cdot \nabla' -{e^2\over r'}+m_ec^2\beta\right)\psi,
\eea
where $\nabla'\equiv (\pa/\pa x'^1)\mathbf{i}+(\pa/\pa x'^2)\mathbf{j}+(\pa/\pa x'^3)\mathbf{k}$ and $r'=\sqrt{(x'^1)^2+(x'^2)^2+(x'^3)^2}$. Then we find out that
\bea\la{1-05}
\alpha'\equiv \alpha_z={e^2\over \hbar c}.
\eea
Comparing (\ref{1-05}) with (\ref{solution1}), we conclude that
\bea\la{1-06}
\alpha_z=\alpha,
\eea
which indicates that the fine-structure constant is unvarying indeed in SM of physics.

The argument above on $\alpha$-unvarying in SM made by comparing Dirac equation for laboratory hydrogen atom with that for a distant hydrogen atom  is deeply related to aspects of Special Relativity (SR), General Relativity (GR) and Cosmology. In other words, the $\alpha$-varying phenomena reported in \cite{Webb1,Webb0,Webb99, Webb992, Webb01, Webb03} implies some new physics beyond SM. Further remarks on this issue are follows:
\begin{enumerate}
\item The spectrum equations (\ref{1-02}) (\ref{1-04}) come from the following inhomogeneous-Lorentz (or Poincar\'{e}) invariant (i.e., $ISO(3,1)$) Dirac equation and Maxwell equation in local inertial systems of FRW Universe:
\bea
\la{Dirac01} &&(i \gamma^\mu D_\mu^L -{m_e c\over \hbar })\psi=0,\\
\la{Max}&& F^{\mu\nu}_{~~~,\;\nu}=j^\mu=-\delta^{\mu 0}4\pi e\delta^{(3)}(\mathbf{x}),
\eea
where $D_\mu^L={\pa\over \pa L^\mu}-ie/(c\hbar)\eta_{\mu\nu}A^\nu$, and the electromagnetic potential $A^\nu\equiv \{\phi=e/r,\;\mathbf{A}\}$. So, the operator structure of (\ref{1-02}), (\ref{1-04}) and a dimensionless combination of universal constants $\alpha=e^2/(\hbar c)$ are rooted in the symmetry assumption of the theory.

\item The point that $\alpha$ is unvarying is deduced from the constancy of the adopted metric of local inertial coordinate system in the FRW Universe (i,e., $\{\eta_{\mu\nu}\}=$const.). So, the fact of the $\alpha$-varying in real world reported in \cite{Webb1,Webb0,Webb99, Webb992, Webb01,Webb03} indicates that the metric of local inertial coordinate system in the real Universe may be spacetime-dependent.
\item  Minkowski metric $\eta_{\mu\nu}=\text{diag}\{+,-,-,-\}$ is the basic spacetime metric of Einstein's Special Relativity (E-SR) in SM. The most general transformation to
preserve metric $\eta_{\mu\nu}$ is Poincar\'e group (or
inhomogeneous Lorentz group $ISO(1,3)$). It is well known that the
Poincar\'e group is the limit of the de Sitter group with pseudo-sphere
radius $|R|\rightarrow \infty$. Therefore, E-SR may possibly be extended to a SR theory with de Sitter space-time symmetry. Since P.A.M. Dirac's work in 1935 \cite{Dirac35} many discussions (e.g., E. In\"{o}n\"{u} and E. P. Wigner in 1968 \cite{Wigner}; F. G\"{u}rsey and T.D. Lee in 1963 \cite{Lee},  etc ) pointed to such a possible extension of E-SR. In 1970's, K.H.Look (Qi-Keng Lu) and his
collaborators Z.L.Zou, H.Y.Guo suggested the de Sitter Invariant Special Relativity (dS-SR) \cite{look}\cite{Lu74} (see Appendix A, and also \cite{Guo1,Ours} and Appendix in \cite{Sun} for the English version).  It has been proved that Lu-Zou-Guo's dS-SR is a satisfying and self-consistent special relativity theory. In 2005, one of us (MLY) and Xiao, Huang, Li suggested dS-SR Quantum Mechanics (QM) \cite{Ours}.

 \item Beltrami metric (see Appendix A)
\bea\la{00}B_{\mu\nu}(x)=\eta_{\mu\nu}/\sigma(x)+\eta_{\mu\lambda}x^\lambda \eta_{\nu\rho}x^\rho/(R^2\sigma(x)^2),~~{\rm with}\;\sigma(x)=1-\eta_{\mu\nu}x^\mu x^\nu/R^2>0
\eea
is the basic metric of dS-SR with Minkowski point coordinates $M^\mu=0$ (i.e., $B_{\mu\nu}(x)|_{x=M=0}$ $=\eta_{\mu\nu}$) \cite{Ours}. Both $\eta_{\mu\nu}$ and $B_{\mu\nu}$ lead to the inertial motion law for free particles $\ddot{\mathbf{x}}=0$, which is the precondition to define inertial reference systems required by special relativity theories. However, the $B_{\mu\nu}$-preserving coordinate transformation group is de Sitter group $SO(4,1)$ (or $SO(3,2)$) rather than E-SR's inhomogeneous Lorentz group $ISO(1,3)$ \cite{look,Lu74,Guo1,Ours}, different from the case of $\eta_{\mu\nu}$. In addition, $\eta_{\mu\nu}$ does not satisfy the Einstein equation with $\Lambda$ (Einstein cosmology constant) in vacuum, but $B_{\mu\nu}(x)$ does satisfy it (see below). Generally, when $M^\mu\neq 0$, the basic metric of dS-SR is modified to be
\bea\la{1-1}
B_{\mu\nu}^{(M)}(x)\equiv B_{\mu\nu}(x-M)={\eta_{\mu\nu}\over \sigma^{(M)}(x)}+{\eta_{\mu\lambda}(x^\lambda-M^\lambda) \eta_{\nu\rho}(x^\rho-M^\rho)\over R^2 \sigma^{(M)}(x)^2},
\eea
where
\bea\la{1-2}
\sigma^{(M)}(x)\equiv\sigma(x-M)=1-{\eta_{\mu\nu}(x^\mu-M^\mu) (x^\nu-M^\nu)\over R^2}.
\eea
which will be called Modified Beltrami metric, or M-Beltrami metric. Based on $B_{\mu\nu}^{(M)}(x)$, the dS-invariant special relativity with $M^\mu\neq 0$ can be built. The procedures and formulation are similar to ordinary dS-SR in \cite{look,Lu74,Guo1,Ours}, which is actually a slight extension of usual dS-SR (see Appendix B).
It is essential, however, that $B_{\mu\nu}^{(M)}(x)$ is spacetime dependent and has more parameters $\{R,\;M^\mu\}$, which may provide a possible clue to solve the puzzle of $\alpha$-varying.

\item Our strategy in this present paper for solving this puzzle is to pursue the
 following dS-SR Dirac equation for both electron in laboratory hydrogen and electron in distant hydrogen in the FRW Universe (see Eq.(25) in \cite{Yan0}):
\bea
\la{Dirac0} (ie_{a}^{~\mu} \gamma^a\mathcal{D}_\mu^L -{m_e c\over \hbar })\psi=0,
\eea
where $\mathcal{D}_\mu^L={\pa\over \pa L^\mu}-{i\over
4}\omega^{ab}_{\;\;\mu}\sigma_{ab}-ie/(c\hbar)B_{\mu\nu}^{(M)}A^\nu$,
$e^{~\mu}_a$ is the tetrad, $\omega_{~~\mu}^{ab}$ is spin-connection, and the electromagnetic potential
$A^\nu\equiv \{\phi_B,\;\mathbf{A}\}$. Unlike E-SR Dirac equation (\ref{Dirac01}), the spacetime symmetry of
(\ref{Dirac0}) is de Sitter invariant group $SO(4,1)$ (or $SO(3,2)$) instead of former $ISO(3,1)$.
Specifically, the dS-SR Dirac spectrum equation can be deduced from (\ref{Dirac0}). It is essential that the result will be different form E-SR equation (\ref{1-02}). Following the method used in (\ref{1-02}) (\ref{solution1}), the coefficients of resulting $(-i\bm{\alpha}\cdot\nabla)$-type and $(-1/r)$-type operator terms in the dS-SR Dirac spectrum equation are of $\hbar_z(\mathbf{\Omega})c$ and $e_z(\mathbf{\Omega})^2$ respectively. Then their ratio yields prediction of $\alpha_z(\mathbf{\Omega})\equiv e_z(\mathbf{\Omega})^2/(\hbar_z(\mathbf{\Omega})c)$. The adjustable parameters in this model are $R$ and the position of Minkowski point $M^\mu$. For simplicity, we take $M^\mu=\{M^0,\;M^1,\;0,\;0\}$. It turns out to be a good a choice for solution to the puzzle of $\alpha$-varying.

\item  Different from Quantum Mechanics (QM) wave equation (\ref{Dirac01}) deduced from $\eta_{\mu\nu}$, the equation (\ref{Dirac0}) is actually a time-dependent Hamiltonian problems in QM. This is because $B_{\mu\nu}^{(M)}(x)$ is time-dependent.Therefore the corresponding Lagrangian $L_{dS}$ (see Eq.(\ref{3-26})) and hence Hamiltonian is time-dependent \cite{Ours}. In this paper, the adiabatic approach \cite{Born}\cite{Messian}\cite{Bayfield} will be used to deal with the time-dependent Hamiltonian problems in dS-SR QM.
Generally, to a $H(x,t)$, we may express it as
$H(x,t)=H_0(x)+H'(x,t)$. Suppose two eigenstates $|s\rangle$ and $|m
\rangle$ of $H_0(x)$ are not degenerate, i.e., $\Delta E\equiv \hbar (
\omega_{m}-\omega_s)\equiv\hbar \omega_{ms}\neq 0$.
 The validity of for  adiabatic
approximation relies on the fact that the variation of the potential
$H'(x,t)$ in the the Bohr time-period $(\Delta
T_{ms}^{(Bohr)})\dot{H}'_{ms}=(2\pi/\omega_{ms})\dot{H}'_{ms}$
is much less than $\hbar \omega_{ms}$, where $H'_{ms}\equiv \langle m|H'(x,t)|s\rangle$. That  makes the quantum
transition from  state $|s\rangle$ to  state $|m \rangle$ almost
impossible. Thus,  the non-adiabatic effect corrections are small
enough (or tiny) , and  the adiabatic approximations are legitimate
. For the wave equation of dS-SR QM of atoms discussed in this paper,
we show that the perturbation Hamiltonian describes the time evolutions of the system $H'(x,t)\propto (c^2t^2/ R^2)$ (where $t$ is the cosmic time). Since $R$ is
cosmologically large and $R>>ct$, the factor
$(c^2t^2/ R^2)$ will make the
time-evolution of the system so slow that the adiabatic
approximation works. We shall  provide a calculations to
 confirm this point in the paper.
By this approach, we solve the
stationary dS-SR Dirac equation for one electron atom, and
the spectra of the corresponding Hamiltonian with time-parameter are
obtained. Consequently, we find out that the electron mass $m_e$, the electric charge $e$, the Planck constant $\hbar$ and the fine structure constant
$\alpha=e^2/(\hbar c)$ vary  as cosmic time goes by.
 These are interesting consequences since they indicate that the time-variations of fundamental physics
constants are due to solid known quantum evolutions of time-dependent
quantum mechanics that has been widely discussed for a long history (e.g., see
\cite{Bayfield} and the references within).

\item Finally, we argue that it is reasonable to assume that the Beltrami metric is the appropriate metric for the spacetime of the local inertial system in real world.  If we express the total energy momentum tensor $T_{\mu\nu}$ as the sum of a possible vacuum term $-\rho_{(v)}g_{\mu\nu}$ and a term $T^M_{\mu\nu}$ arising from matter (including radiation), then the complete Einstein equation is \cite{Peebles}\cite{Pad}\cite{YanH}:
\bea\la{5-75}
\mathcal{R}_{\mu\nu}-{1\over2}g_{\mu\nu}\mathcal{R}+\Lambda g_{\mu\nu}=-{8\pi G}T_{\mu\nu}^M -{8\pi G}\rho_{(v)}g_{\mu\nu},
\eea
where $\rho_{(v)}$ is the dark energy density, so $\Lambda_{\rm dark \;energy}= 8\pi G\rho_{(v)}$, and $\Lambda$ is originally introduced by Einstein in 1917, and serves as a universal constant in physics. We call it the Einstein (or geometry) cosmological constant. The effective cosmologic constant $\Lambda_{eff}=\Lambda+\Lambda_{\rm dark\; energy}\simeq 1.26 \times 10^{-56}\; {\rm cm}^{-2}$ is the observed value determined via effects of accelerated expansion of the universe \cite{L1} and the recent WMAP data \cite{L2}. We have no any {\it a priori} reason to assume the geometry cosmologic constant to be zero, so the vacuum Einstein equation is:
\bea\la{13}
\mathcal{R}_{\mu\nu}-{1\over2}g_{\mu\nu}\mathcal{R}+\Lambda g_{\mu\nu}=0,
\eea
instead of $G_{\mu\nu}=0$, and hence the vacuum solution to (\ref{13}) is $g_{\mu\nu}=B^{(M)}_{\mu\nu}(x)$ with $|R|=\sqrt{3/ \Lambda}$ instead of $g_{\mu\nu}=\eta_{\mu\nu}$. Therefore, we conclude that the metric of the local inertial coordinate system in real world should be Beltrami metric rather than Minkowski metric. Thus, the dS-SR Dirac equation (\ref{Dirac0}) (instead of E-SR Dirac equation (\ref{Dirac01})) is  legitimate to characterize the spectra in the real world, and then the $\alpha$-varying over the real world space-time would occur naturally.
\end{enumerate}

\noindent  This paper provides an understanding of the $\alpha$-varying in cosmology reported in \cite{Webb1} \cite{Webb0} by means of extending the basic spacetime symmetry in the local inertial coordinate systems in the standard cosmologic model. The contents of the paper are organized as follows. In section 2, light-cone of Friedmann-Robertson-Walker universe is described, and relation of cosmological time to redshift $z$ is shown. The relation of $t-z$ is based on $\Lambda$CDM model and the cosmology parameters $(H_0,\;\Omega_{m0},\; \Omega_\Lambda\simeq 1-\Omega_{m0})$ in real world; In section 3, we describe the local inertial coordinate system in light-cone of FRW Universe with Einstein cosmology constant $\Lambda$. The metric of such local inertial systems is M-Beltrami metric that services as the basic metric of de Sitter invariant special relativity (dS-SR); In section 4, we derive the electric Coulomb law at light-cone of FRW Universe in terms of dS-SR Maxwell equations. As is well known, Coulomb force dominates the dynamics of the atomic spectrums. In section 5, we discuss the fine-structure constant variation along the best-fit dipole direction shown in \cite{Webb1,Webb0}. The $\alpha$-varying $\Delta\alpha/\alpha_0\equiv (\alpha_z(\Omega)-\alpha_0)/\alpha_0$ (where $\alpha_0$ is $\alpha$'s value in laboratory) in this region is derived. Using the data along this best-fit dipole reported by \cite{Webb1,Webb0}, the model's parameters are determined. The theoretical predictions are consistent with the observations. In section 6, we examine the $\alpha$-varying in whole sky. The results are also in agreement with the estimate from Keck- and VLT data. Finally, we briefly summarize and discuss our results. In Appendix A we briefly recall the Betrami metric and the de Sitter invariant special relativity. In Appendix B, a remark on the modified Beltrami metric used in this paper is provided.

\section{Light-Cone of Friedmann-Robertson-Walker Universe}

\begin{figure}[h]
\begin{center}
\includegraphics[scale=0.4]{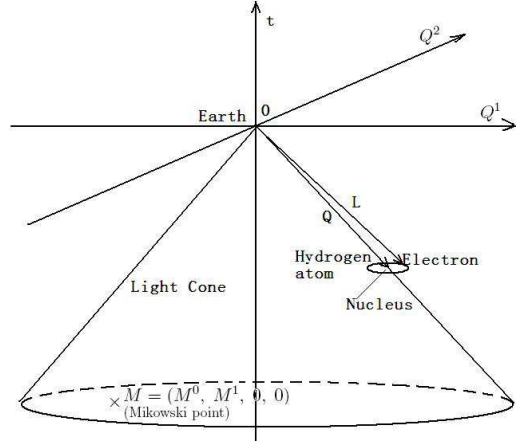}
\caption{\label{Fig1} \footnotesize Sketch of the light cone of the Friedmann-Robertson-Walker Universe. Only 3 coordinate axes $\{Q^0=ct,\;Q^1,\;Q^2 \}$ are shown in this three dimensional figure. The $Q^3$ axis could be imagined. The Earth is located in the origin. The position vector for
nucleus of atom between the QSO and the Earth is $\mathbf{Q}$, and for electron is $\mathbf{L}$. The distance
between nucleus and electron is $\bar{r}\sim |\mathbf{L}-\mathbf{Q}|$. The location of the Minkowski point of Betrami metric is denoted by notation $``\times"$ with $M=(M^0,\;M^1,\;0,\;0)$. }
\end{center}
\end{figure}

\noindent The isotropic and homogeneous cosmology solution of Einstein equation in GR (General Relativity) is Friedmann-Robertson-Walker (FRW) metric. In this section we discuss the Light-Cone of FRW Universe because all visible quasars in sky must be located on it (see Figure \ref{Fig1}).

\begin{figure}[h]
\begin{center}
\includegraphics[scale=0.6]{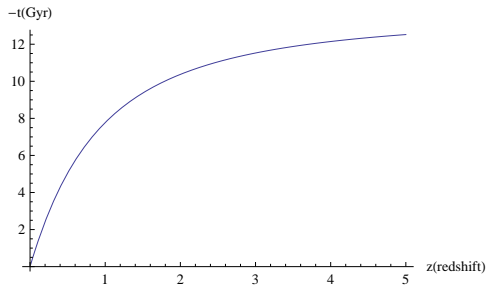}
\caption{\label{fig2}The $t-z$ relation in $\Lambda$CDM model
(eq.(\ref{newLa1})).}
\end{center}
\end{figure}

The
Friedmann-Robertson-Walker (FRW) metric is (see, e.g., \cite{Weinberg})
\begin{eqnarray}
\nonumber ds^2&=&c^2dt^2-a(t)^2\left\{{dr^2\over
1-kr^2}+r^2d\theta^2
+r^2\sin^2\theta d\phi^2\right\} \\
\nonumber&=&(dQ^0)^2-a(t)^2\left\{dQ^idQ^i+{k (Q^idQ^i)^2\over
1-kQ^iQ^i}\right\} \\
\label{new4}&\equiv& g_{\mu\nu}(Q) dQ^\mu dQ^\nu,
\end{eqnarray}
where $a(t)$ is scale (or expansion) factor and  $r=\sqrt{Q^iQ^i}\equiv Q,~Q^1=Q\sin \theta \cos \phi,~Q^2=Q\sin \theta
\sin \phi,~Q^3=Q\cos\theta$ and $(Q^0)^2=c^2t^2$. Hereafter, for the sake of convenience, we take $t$ to be looking-back cosmologic time, so that $t<0$. As is well know FRW
metric satisfies ¡°homogeneity and isotropy¡± principle of present
day cosmology.
For simplicity, we take $k=0$ and $a(t)=1/(1+z(t))$ (i.e., $a(t_0)=1$). And the red
shift function $z$ is determined by $\Lambda$CDM model
 \cite{Lambda,Peebles,Lambda1}(see, e.g., Eq.(64) of \cite{Peebles}):
\begin{equation}\label{newLa1}
t(z)=\int_z^0{dz' \over H(z')(1+z')},
\end{equation}
where
\begin{eqnarray}
\nonumber \label{newLa2}H(z')&=&H_0\sqrt{\Omega_{m0}(1+z')^3+\Omega_{R0}(1+z')^4+1-\Omega_{m0}},\\
\nonumber \label{newLa3}H_0&=&100\;h\simeq 100\times0.705\, {\rm km}\cdot {\rm s}^{-1}/{\rm Mpc},\\
\label{newLa4}\Omega_{m0}&\simeq &0.274,~~~\Omega_{R0}\sim 10^{-5}.
\end{eqnarray}
Figure of $t(z)$ of Eq.(\ref{newLa1}) is shown in Figure \ref{fig2}.

\begin{figure}[h]
\begin{center}
\includegraphics[scale=0.6]{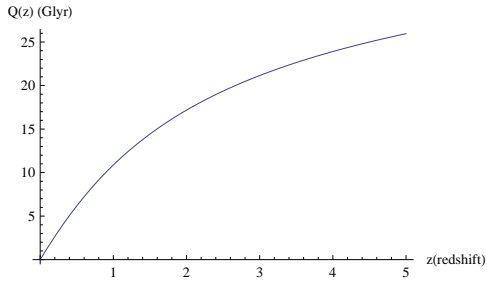}
\caption{\label{fig3}Function $Q(z)$ in $\Lambda$CDM model
(eq.(\ref{5})).}
\end{center}
\end{figure}

\begin{figure}[h]
\begin{center}
\includegraphics[scale=0.6]{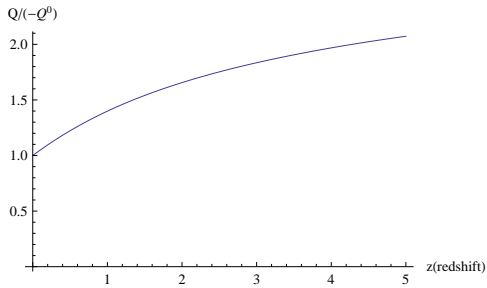}
\caption{\label{R}Function of $Q(z)/Q^0(z)$. $Q(z)$ and $Q^0(z)=ct$ are given in Eqs. (\ref{5}) and (\ref{newLa1}).}
\end{center}
\end{figure}

The Light-Cone of FRW Universe is defined by $ds^2=0$. From Eq.(\ref{new4}), we have the light-cone equation:
\bea\la{4}
(dQ^0)^2-a(t)^2(dQ)^2=0,~~~{\rm or}~~~-cdt=a(t)dQ={1\over 1+z(t)}dQ.
\eea
Substituting (\ref{newLa1}) into (\ref{4}) gives
\bea\la{5}
Q(z)=c\int_0^z{dz'\over H(z')}.
\eea
Figure of $Q(z)$ of Eq.(\ref{5}) is shown in figure \ref{fig3}. Ratio of $Q$ over $Q^0$ is shown in figure \ref{R}.

\section{Local Inertial Coordinate System in Light-Cone of FRW Universe with Einstein Cosmology Constant }
In principle, almost all calculations on quantum spectrums in atomic physics are achieved in the inertial coordinate systems. From the cosmological point of view, the phenomena of atomic spectrums should be described in the local inertial coordinate systems of FRW Universe. Therefore, we are interested in how to determine the local inertial coordinate system in light-cone of FRW Universe when the Einstein cosmology constant $\Lambda$ is present.

Existence of {\it local inertial coordinate system} is required by the Equivalence Principle. The principle states that experiments in a sufficiently small falling laboratory, over a sufficiently short time, give results that are indistinguishable from those of the same experiments in an inertial frame in empty space of special relativity \cite{Hartle}. Such a sufficiently small falling laboratory, over a sufficiently short time represents {\it a local inertial coordinates system.}  This principle suggests that the local properties of curved spacetime should be indistinguishable from those of the spacetime with {\it inertial metric} of special relativity. A concrete expression of this ideal is the requirement that, given a metric $g_{\alpha\beta}$ in one system of coordinates $x^\alpha$, at each point $P$ of spacetime
it is possible to introduce new coordinates $x'^\alpha$ such that
\begin{equation}\label{bb1}
g'_{\alpha\beta}(x'_P)={\rm inertial\;metric\;of\;SR\;at\;} x'_P,
\end{equation}
and the connection at $x'_P$ is the Christoffel symbols deduced from $g'_{\alpha\beta}(x'_P)$.

In usual Einstein's general relativity (without $\Lambda$), the above expression is
\begin{equation}\label{bb2}
g'_{\alpha\beta}(x'_P)=\eta_{\alpha\beta},~~and~~\Gamma^\lambda_{\alpha\beta}=0,
\end{equation}
which satisfies the Einstein equation of E-GR in empty space: $G_{\mu\nu}=0$.

In dS-GR (GR with a $\Lambda$), the local inertial coordinate system at $x'^\alpha_P$ is characterized by
\begin{eqnarray}\label{bb3}
g'_{\alpha\beta}(x'_P)&=&
B_{\alpha\beta}^{(M)}(x'_P)\equiv {\eta_{\mu\nu}\over \sigma^{(M)}(x'_P)}+{\eta_{\mu\lambda}(x'^\lambda-M^\lambda) \eta_{\nu\rho}(x'^\rho_P-M^\rho)\over R^2 \sigma^{(M)}(x'_P)^2},\\
\nn && {\rm with} \;\;
\sigma^{(M)}(x'_P)=1-{\eta_{\mu\nu}(x'^\mu_P-M^\mu) (x'^\nu_P-M^\nu)\over R^2},\\
\nn \Gamma^\lambda_{\alpha\beta}&=& \frac{1}{ 2}(B^{(M)})^{\lambda\rho}
(\pa_\alpha B_{\rho\beta}^{(M)}+\pa_\beta B_{\rho\alpha}^{(M)}-\pa_\rho B_{\alpha\beta}^{(M)})\\
\la{bb4}&=&{1\over R^2\sigma^{(M)}(x'_P)}(\delta_\mu^\lambda \eta_{\nu\rho}+\delta_\nu^\lambda \eta_{\mu\rho})(x'^\rho_P-M^\rho),\\
\nn && \hskip-1in ( {\rm or}\;\;  {\pa g'_{\rho\beta}(x'_P)\over \pa x'^\alpha_P} ={x'^\nu_P-M^\nu\over R^2\sigma^{(M)}(x'_P)^2}
\left[ 2\eta_{\rho\beta}\eta_{\alpha\nu}+\eta_{\rho\alpha}\eta_{\beta\nu}+\eta_{\beta\alpha}\eta_{\rho\nu}
+{4\eta_{\rho\mu}\eta_{\beta\nu}\eta_{\alpha\lambda}(x'^\mu_P-M^\mu)( x'^\lambda_P-M^\lambda)\over R^2\sigma^{(M)}(x'_P)}\right] )
\end{eqnarray}
where $B_{\alpha\beta}^{(M)}(x'_P)$ were given in (\ref{1-1}), which satisfies the Einstein equation of dS-GR in empty spacetime: $G_{\mu\nu}+\Lambda g_{\mu\nu}=0$ with $\Lambda=3/R^2$.
 (Note $\eta_{\mu\nu}$ does not satisfy that equation, i.e., $G_{\mu\nu}(\eta)+\Lambda \eta_{\mu\nu}\neq 0$. So it cannot be the metric of the local inertial system in dS-GR with $\Lambda$).

To the light cone of FRW Universe with $\Lambda$, the coordinate-components $Q^0(z)=ct(z)$ and $Q(z)$ have been shown in Eqs.(\ref{newLa1}) (or Figure \ref{fig2}) and (\ref{5}) (or Figure \ref{fig3}) respectively. Therefore from (\ref{1-1}), the space-time metric of the local inertial coordinate system at position $Q(z)$ of the light cone is determined to be
\bea\la{01-1}
B_{\mu\nu}^{(M)}(\mathcal{Q})\equiv B_{\mu\nu}(\mathcal{Q}-M)={\eta_{\mu\nu}\over \sigma^{(M)}(\mathcal{Q})}+{\eta_{\mu\lambda}(Q^\lambda-M^\lambda) \eta_{\nu\rho}(Q^\rho-M^\rho)\over R^2 \sigma^{(M)}(\mathcal{Q})^2},
\eea
where
\bea\la{01-2}
\sigma^{(M)}(\mathcal{Q})\equiv\sigma(\mathcal{Q}-M)=1-{\eta_{\mu\nu}(Q^\mu-M^\mu) (Q^\nu-M^\nu)\over R^2}.
\eea
We see from Figure \ref{Fig1} that the visible atom is embedded into the light cone at $Q$-point. Since $Q\simeq L$ (i.e., comparing with the Universe, atoms are very very small), we can reasonably treat the metric of the spacetime in the atomic region as a constant. This is just the adiabatic approximation adopted in \cite{Yan0}. When $Q^\mu\rightarrow M^\mu$, we have $B_{\mu\nu}^{(M)}(\mathcal{Q})\Rightarrow \eta_{\mu\nu}$. So, $Q^\mu=M^\mu$ is the Minkowski point of the Beltromi metric $B_{\mu\nu}^{(M)}(\mathcal{Q})$.

Now let's derive $e^{~\mu}_a$ and $\omega^{ab}_{\;\;\mu}$ from (\ref{01-1}). Setting
\bea\la{01-3}
q^\mu\equiv Q^\mu-M^\mu,
\eea
then equations (\ref{01-1}) and (\ref{01-2}) become
\bea\la{01-4}
B_{\mu\nu}^{(M)}={\eta_{\mu\nu}\over \sigma^{(M)}}+{\eta_{\mu\lambda}q^\lambda \eta_{\nu\rho}q^\rho\over R^2 (\sigma^{(M)})^2},
\eea
where
\bea\la{01-5}
\sigma^{(M)}\equiv\sigma^{(M)}(q)=1-{\eta_{\mu\nu}q^\mu q^\nu\over R^2}.
\eea
We introduce notations:
\bea\la{6-41}
&&\bar{q}_\mu\equiv \eta_{\mu\lambda}q^\lambda,~~~q^\lambda\equiv \bar{q}^\lambda=\eta^{\mu\lambda}\bar{q}_\mu,\\
&&\bar{q}^2\equiv \eta_{\mu\nu}q^\mu q^\nu=\bar{q}_\nu \bar{q}^\nu,
\eea
and construct
two project operators in spacetime $\{\bar{q}^\mu\}$ with metric $\eta_{\mu\nu}$:
\bea\la{6-44}
\bar{\theta}_{\mu\nu}\equiv\eta_{\mu\nu}-{\bar{q}_\mu \bar{q}_\nu\over \bar{q}^2},~~~~\bar{\omega}_{\mu\nu}\equiv{\bar{q}_\mu \bar{q}_\nu\over \bar{q}^2}.
\eea
It is easy to check the calculation rules for project operators:
\bea\la{6-45}
&& \bar{\theta}_{\mu\lambda}\bar{\theta}^\lambda_{~~\nu}\equiv \bar{\theta}_{\mu\lambda}\eta^{\lambda\rho}\bar{\theta}_{\rho\nu} =\bar{\theta}_{\mu\nu},~~~{\rm or\;in\;short}~\bar{\theta}\cdot\bar{\theta}=\bar{\theta},\\
\la{6-46}&& \bar{\omega}_{\mu\lambda}\bar{\omega}^\lambda_{~~\nu} \equiv \bar{\omega}_{\mu\lambda}\eta^{\lambda\rho}\bar{\omega}_{\rho\nu}=\bar{\omega}_{\mu\nu},~~~{\rm or\;in\;short}~~~\bar{\omega}\cdot\bar{\omega}=\bar{\omega},\\
\la{6-47}&& \bar{\theta}_{\mu\lambda}\bar{\omega}^\lambda_{~~\nu} \equiv \bar{\theta}_{\mu\lambda}\eta^{\lambda\rho}\bar{\omega}_{\rho\nu}=0,~~~{\rm or\;in\;short}~~~\bar{\theta}\cdot\bar{\omega}=0,\\
\la{6-48}&& \bar{\theta}_{\mu\nu}+ \bar{\omega}_{\mu\nu}=\eta_{\mu\nu},~~~{\rm or\;in\;short}~~~\bar{\theta}+\bar{\omega}=I.
\eea
$B_{\mu\nu}^{(M)}(Q)$ can be written as
\bea
B_{\mu\nu}^{(M)}
\la{6-40}= {\eta_{\mu\nu}\over \sigma^{(M)}}+{\bar{q}_\mu\bar{q}_\nu\over R^2(\sigma^{(M)})^2},
\eea
where
\bea
\la{6-42} \sigma^{(M)}=1-{\eta_{\mu\nu}q^\mu q^\nu\over R^2}=1-{\bar{q}_\nu \bar{q}^\nu\over R^2}\equiv 1-{\bar{q}^2\over R^2}.
\eea
Since $B_{\mu\nu}^{(M)}$ is a tensor in the spacetime $\{\bar{q}^\mu,~\eta_{\mu\nu}\}$,  it can be written as follows from (\ref{6-40}):
\bea\la{6-49}
B_{\mu\nu}^{(M)}={1\over\sigma^{(M)}}\bar{\theta}_{\mu\nu}+{1\over(\sigma^{(M)})^2}\bar{\omega}_{\mu\nu}.
\eea
Furthermore, by means of $(B^{(M)})^{\mu\nu}B^{(M)}_{\nu\lambda}=\delta^\mu_\nu\equiv \eta^\mu_\nu$ and the rules (\ref{6-44})-(\ref{6-48}), the above expression of $B_{\mu\nu}^{(M)}$ leads to:
\bea\la{6-50}
(B^{(M)})^{\mu\nu}=\sigma^{(M)} \bar{\theta}^{\mu\nu}+(\sigma^{(M)})^2\bar{\omega}^{\mu\nu}.
\eea
Or explicitly in matrix form:
\bea\nn
\left\{(B^{(M)})^{\mu\nu}\right\}=\left(
\begin{array}{lccr}
\sigma^{(M)}(1-{(q^0)^2\over R^2}) & {-q^0q^1\sigma^{(M)}\over R^2} &{-q^0q^2\sigma^{(M)}\over R^2}
 &{-q^0q^3\sigma^{(M)}\over R^2} \\
{-q^1q^0\sigma^{(M)}\over R^2} & -\sigma^{(M)}(1+{(q^1)^2\over R^2}) & {-q^1q^2\sigma^{(M)}\over R^2}
 & {-q^1q^3\sigma^{(M)}\over R^2} \\
{-q^2q^0\sigma^{(M)}\over R^2} & {-q^2q^1\sigma^{(M)}\over R^2} & -\sigma^{(M)}(1+{(q^2)^2\over R^2})
& {-q^2q^3\sigma^{(M)}\over R^2} \\
{-q^3q^0\sigma^{(M)}\over R^2} & {-q^3q^1\sigma^{(M)}\over R^2} & {-q^3q^2\sigma^{(M)}\over R^2}
& -\sigma^{(M)}(1+{(q^3)^2\over R^2})
\end{array} \right)\\
\la{AA2}
\eea
In the Beltrami spacetime $\mathcal{B}$ with the metric $B_{\mu\nu}^{(M)}$, the tetrad $e^a_{~\mu}$ is defined via the following  equation
\bea\la{6-51}
B_{\mu\nu}^{(M)}=\eta_{ab}e^a_{~\mu}e^b_{~\nu}.
\eea
Generally, we have expansion of $e^a_{~\mu}$:
\bea\la{6-52}
e^a_{~\mu}=a\bar{\theta}^a_{~~\mu}+b\bar{\omega}^a_{~~\mu},
\eea
where $a$ and $b$ are unknown constants. Substituting equations (\ref{6-49}) (\ref{6-52}) into (\ref{6-51}) gives
\bea\nn
{1\over\sigma^{(M)}}\bar{\theta}_{\mu\nu}+{1\over(\sigma^{(M)})^2}\bar{\omega}_{\mu\nu}&=&\eta_{ab}
(a\bar{\theta}^a_{~~\mu}+b\bar{\omega}^a_{~~\mu})(a\bar{\theta}^b_{~~\nu}+b\bar{\omega}^b_{~~\nu})\\
\la{6-53}&=&a^2\bar{\theta}_{\mu\nu}+b^2\bar{\omega}_{\mu\nu}.
\eea
Comparing the left side of (\ref{6-53}) with the right side, and noting $\bar{\theta}$ and $\bar{\omega}$ being project operators with properties of (\ref{6-45})-(\ref{6-48}), we find that:
\bea\la{6-54}
a=\sqrt{1\over \sigma^{(M)}},~~~~b={1\over\sigma^{(M)}}.
\eea
Substituting (\ref{6-54}) into (\ref{6-52}) gives
\bea\nn
e^a_{~\mu}&=&\sqrt{1\over\sigma^{(M)}}\bar{\theta}^a_{~~\mu}+{1\over\sigma^{(M)}}\bar{\omega}^a_{~~\mu}\\
\la{6-55}&=&\sqrt{1\over\sigma^{(M)}}\delta^a_{\mu}+\left({1\over\sigma^{(M)}}-{1\over\sqrt{\sigma}^{(M)}}
\right) {\eta_{\mu\nu}\delta^a_\lambda(Q^\lambda-M^\lambda) (Q^\nu-M^\nu)\over(1-\sigma^{(M)})R^2}.
\eea
$e_a^{~\mu}$ is the inverse of $e^a_{~\mu}$ given by:
\bea\la{6-19+}
e_a^{~\mu}e^{a'}_{~\mu}=\delta^{a'}_a.
\eea
With the equations (\ref{6-19+}) and (\ref{6-51}), we have
\bea\nn
e^{~\mu}_a&=&e^b_{~\nu}(B^{(M)})^{\nu\mu}\eta_{ba}=
(\sqrt{1\over\sigma^{(M)}}\bar{\theta}^b_{~~\nu}+{1\over\sigma^{(M)}}\bar{\omega}^b_{~~\nu})
(\sigma^{(M)}\bar{\theta}^{\nu\mu}+(\sigma^{(M)})^2\bar{\omega}^{\nu\mu})\eta_{ab}\\
\nn&=&\sqrt{\sigma^{(M)}}\bar{\theta}_a^{~\mu}+\sigma^{(M)}\bar{\omega}_a^{~\mu}\\
\la{6-57}&=&\sqrt{\sigma^{(M)}}\delta_a^{\mu}+{\sigma^{(M)}-\sqrt{\sigma^{(M)}}\over 1-\sigma^{(M)}}{\eta_{ab}\delta^b_\lambda(Q^\lambda-M^\lambda)(Q^\mu-M^\mu)\over R^2},
\eea
and
\bea\nn
e^{a\mu}&=&e^a_{~\nu}(B^{(M)})^{\nu\mu}=(\sqrt{1\over\sigma^{(M)}}\bar{\theta}^a_{~~\nu}+{1\over\sigma^{(M)}} \bar{\omega}^a_{~~\nu})
(\sigma^{(M)}\bar{\theta}^{\nu\mu}+(\sigma^{(M)})^2\bar{\omega}^{\nu\mu})\\
\nn&=&\sqrt{\sigma^{(M)}}\bar{\theta}^{a\mu}+\sigma^{(M)}\bar{\omega}^{a\mu}\\
\la{6-58}&=&\sqrt{\sigma^{(M)}}\delta^\mu_b\eta^{ab}+{\sigma^{(M)}-\sqrt{\sigma^{(M)}}\over 1
-\sigma}{\delta^a_\mu(Q^\mu-M^\mu) (Q^\mu-M^\mu)\over R^2},\\
\nn
e_{a\mu}&=&\sqrt{1\over\sigma^{(M)}}\bar{\theta}_{a\mu}+{1\over\sigma^{(M)}}\bar{\omega}_{a\mu}\\
\la{6-59}&=&\sqrt{1\over\sigma^{(M)}}\delta^\nu_a\eta_{\nu\mu}+\left({1\over\sigma^{(M)}}-{1\over\sqrt{\sigma^{(M)}}}
\right) {\eta_{\mu\nu}\eta_{ab}\delta^b_\lambda(Q^\lambda-M^\lambda) (Q^\nu-M^\nu)\over(1-\sigma^{(M)})R^2}.
\eea

 Next we derive spin-connection $\omega^{ab}_{~~\mu}$. From identity
\bea\la{2-2}
&&e^\mu_{a\; ;\nu}=\pa_\nu e_a^\mu+ \omega_{a\;\;\nu}^{\;\;b}e_b^\mu+\Gamma^\mu_{\lambda\nu} e^\lambda_a=0,\\
\la{2-3}&& \Gamma^\rho_{\lambda\mu} = {1\over
2}(B^{(M)})^{\rho\nu}(\pa_\lambda B^{(M)}_{\nu\mu} +\pa_\mu B^{(M)}_{\nu\lambda}
-\pa_\nu B^{(M)}_{\lambda\mu}),
\eea
we have
\begin{eqnarray}
\la{2-4} \omega^{ab}_{~~\mu} = {1\over 2}(e^{a\rho}\pa_\mu e^b_\rho
-e^{b\rho}\pa_\mu e^a_\rho ) -{1\over
2}\Gamma^\rho_{\lambda\mu}(e^{a\lambda}e^b_\rho
-e^{b\lambda}e^a_\rho ).
\end{eqnarray}
Substituting (\ref{01-4}) (\ref{6-55}) into (\ref{2-3}) and (\ref{2-4}) gives
\begin{eqnarray}
\label{2-5} \omega^{ab}_{~~\mu}={1\over R^2\left(1+\sqrt{\sigma^{(M)}}\right)\sqrt{\sigma^{(M)}}}(\delta^a_\mu \delta^b_\lambda-\delta^b_\mu \delta^a_\lambda)(Q^\lambda-M^\lambda).
\end{eqnarray}

\section{ Electric Coulomb Law at Light-Cone of FRW Universe}

\noindent The hydrogen atom is a bound state of proton and electron. The electric Coulomb potential binds them together.
The action for deriving
that potential of proton located at
$\mathcal{Q}\equiv Q^\mu=\{Q^0=ct,\;Q^1,\;Q^2,\;Q^3\}$ with background
space-time metric $g_{\mu\nu}\equiv B_{\mu\nu}^{(M)}(\mathcal{Q})$ of
eq.(\ref{01-1}) ( see Fig.\ref{Fig1}) in the Gaussian system of units reads
\begin{equation}\label{action1}
S=-{1\over 16\pi c}\int d^{4}L \sqrt{-g}F_{\mu\nu}F^{\mu\nu}-{e\over
c}\int d^{4}L \sqrt{-g}j^\mu A_\mu,
\end{equation}
where $g=\det (B_{\mu\nu}^{(M)})$, $F_{\mu\nu}={\pa A_\nu\over \pa
L^\mu}-{\pa A_\mu\over \pa L^\nu}$ and
$j^\mu\equiv\{j^0=c\rho_{\rm proton}/\sqrt{B_{00}^{(M)}},\;\mathbf{j}\}$ is the
4-current density vector of proton (see, e.g, Ref.\cite{Landau}:
{\it Chapter 4; Chapter 10, Eq.(90.3)}).
Making space-time variable change of $L^\mu\rightarrow
L^\mu-Q^\mu\equiv x^\mu=\{x^0=ct_L-ct,\;x^i=L^i-Q^i\}$ and noting $L^\mu\simeq Q^\mu$, we have
action $S$ as
\begin{eqnarray}
\nonumber S=-{1\over 16\pi c}&& \hskip-0.1in \int d^{4}x
\sqrt{-\det(B_{\mu\nu}^{(M)}(\mathcal{Q}))}
F_{\mu\nu}F^{\mu\nu}-{e\over c}\int d^{4}x \sqrt{-\det(B_{\mu\nu}^{(M)}(\mathcal{Q}))} j^\mu A_\mu \\
\nn = \left(-{1\over 16\pi c}\right.&& \hskip-0.3in
B_{\mu\lambda}^{(M)}(\mathcal{Q})B_{\nu\rho}^{(M)}(\mathcal{Q}) \int  d^{4}x
F^{\mu\nu}(x)F^{\lambda\rho}(x)
  \left. -{e\over c}\int d^{4}x j^\mu (x) A_\mu (x)
\right)\sqrt{-\det(B_{\mu\nu}^{(M)}(\mathcal{Q}))},\\
\label{action2}
\end{eqnarray}
and the equation of motion $ \delta S/\delta A_\mu(x)=0$ as follows
(see, e.g, \cite{Landau}, Eq. (90.6), {\it pp257})
\begin{equation}\label{action3}
\pa_\nu F^{\mu\nu}=(B^{(M)})^{\nu\lambda}\pa_\nu
F^{\mu}_{~~\lambda}=-{4\pi\over c}j^\mu.
\end{equation}
In Beltrami space, $A^\mu=\{\phi_B,\;\mathbf{A}\}$ (see, e.g.,
\cite{Landau}, eq.(16.2) in {\it pp. 45}) and 4-charge current
$j^\mu=\{c\rho_{\rm proton}/\sqrt{B_{00}^{(M)}}, \;\mathbf{j}\}$. According to
the expression of charge density in curved space in Ref.
\cite{Landau}, ({\it pp.256, Eq. (90.4)}), $\rho_{\rm proton}\equiv
\rho_B={e\over \sqrt{\gamma}}\delta^{(3)}(\mathbf{x})$ and $
\mathbf{j}=0$, where
\begin{eqnarray}\label{L1}
&&\gamma=\det (\gamma_{ij}),\\
\label{L2}&&dl^2=\gamma_{ij}dx^idx^j=\left(-B_{ij}^{(M)}+{B_{0i}^{(M)}B_{j0}^{(M)}\over
B_{00}^{(M)}}\right)dx^idx^j~~~~(see~eq.(84.7)~in~Ref.[19])
\end{eqnarray}
\begin{enumerate}
\item When $\mu=i$ ($i=1,2,3$) in Eq.(\ref{action3}), we have
\begin{equation}\label{AB1} \pa^i\pa_\mu
A^\mu-(B^{(M)})^{\mu\nu}\pa_\mu\pa_\nu A^i=-{4\pi\over c}j^i= 0.
\end{equation}
By means of the gauge condition
\begin{equation}\label{AB2}
\pa_\mu A^\mu=0,
\end{equation}
we have
\begin{equation}\label{AB3} (B^{(M)})^{\mu\nu}\pa_\mu\pa_\nu A^i=0.
\end{equation}
Then
\begin{equation}\label{AB4}  A^i=0
\end{equation}
is a solution that satisfies the gauge condition (\ref{AB2}) (noting
$\pa_0A^0={\pa\over \pa x^0}\phi_B(r_B)=0$ due to ${\pa Q^0\over \pa
x^0}={\pa Q^0\over \pa L^0}=0$ ). Eq.(\ref{AB4}) is the vector
potential.

\item  When $\mu=0$ in Eq.(\ref{action3}), we have the Coulomb's law:
\begin{eqnarray}
\nonumber
-(B^{(M)})^{ij}(\mathcal{Q})\pa_i\pa_j\phi_B(x)&=& -{4\pi\over c} j^0= -{4\pi\over c}{c\rho_B\over \sqrt{B_{00}^{(M)}}}={-4\pi e\over
\sqrt{B_{00}^{(M)}\gamma}}\delta^{(3)}(\mathbf{x})\\
\label{A-1} &=&{-4\pi e\over
\sqrt{-\det(B_{\mu\nu}^{(M)}(\mathcal{Q}))}}\delta^{(3)}(\mathbf{x}),
\end{eqnarray}
where $ B_{00}^{(M)}\gamma=-\det(B_{\mu\nu}^{(M)})$ has been used, and
$B_{\mu\nu}^{(M)}$ (and $(B^{(M)})^{\ij}$) were given in (\ref{01-4}) (and (\ref{AA2})), i.e.,
\bea\la{AAA2}
\left\{(B^{(M)})^{ij}\right\}=\left(
\begin{array}{lcr}
  -\sigma^{(M)}(1+{(q^1)^2\over R^2}) & {-q^1q^2\sigma^{(M)}\over R^2}
 & {-q^1q^3\sigma^{(M)}\over R^2} \\
 {-q^2q^1\sigma^{(M)}\over R^2} & -\sigma^{(M)}(1+{(q^2)^2\over R^2})
& {-q^2q^3\sigma^{(M)}\over R^2} \\
 {-q^3q^1\sigma^{(M)}\over R^2} & {-q^3q^2\sigma^{(M)}\over R^2}
& -\sigma^{(M)}(1+{(q^3)^2\over R^2})
\end{array} \right).
\eea
The equation of Coulomb law (\ref{A-1}) can be compactly rewritten as:
\bea\la{A-54}
-\left(\nabla^T_x\mathfrak{B}^{(M)} \nabla_x\right)\phi_B=-\eta\mathfrak{I},
\eea
where tensor $\mathfrak{B}^{(M)}:= \left\{(B^{(M)})^{ij}\right\}$, operator $\nabla:=\{\pa_i\}$,   $\eta={4\pi e\over
\sqrt{\det(B_{\mu\nu}^{(M)}(\mathcal{Q}))}}$ and $\mathfrak{I}:=\delta^{(3)}(\mathbf{x})$.
Symmetric matrix $\mathfrak{B}^{(M)}$ can be diagonalized via similitude transformation due to matrix $P$:
\bea\la{A1-55}
P^{T}\mathfrak{B}^{(M)}P=\Lambda_B=\left(
\begin{array}{lcr}
 \lambda_1 &0&0 \\
 0 & \lambda_2 & 0 \\
0 & 0 & \lambda_3
\end{array} \right).
\eea
Here, $\lambda_i$ with $i=1,\;2,\;3$ and matrix $P$ can be found in $\det (\mathfrak{B}^{(M)}-\lambda {\mathbf I})=0$. From (\ref{AAA2}), the results are:
\bea\la{A-56}
&&\lambda_1=-\sigma^{(M)}\frac{({\mathbf q})^2+R^2 }{R^2},~~\lambda_2=\lambda_3=-\sigma^{(M)},\\
\la{A-57} && P=\left(
\begin{array}{ccc}
 \frac{{q^1\sqrt{(q^3)^2}}}{q^3 \sqrt{\mathbf{q}^2}} & -\frac{{q^1} {q^2}}{ \sqrt{\mathbf{q}^2[(q^1)^2+(q^3)^2]}} & -\frac{{q^3}}{ \sqrt{(q^1)^2+(q^3)^2}} \\
 \frac{{q^2\sqrt{(q^3)^2}}}{q^3 \sqrt{\mathbf{q}^2}} & \frac{\sqrt{(q^1)^2+(q^3)^2}}{\sqrt{\mathbf{q}^2}} & 0 \\
 \frac{\sqrt{(q^3)^2}}{ \sqrt{\mathbf{q}^2}} & -\frac{{q^3} {q^2}}{ \sqrt{\mathbf{q}^2[(q^1)^2+(q^3)^2]}} & \frac{\sqrt{(q^1)^2}}{\sqrt{(q^1)^2+(q^3)^2}}
\end{array}
\right)
\eea
where $\text{q}^i=Q^i-M^i$ with $i=\{1,\;2,\;3\}$ (see Eq.(\ref{01-3})), and $\mathbf{q}^2=(\text{q}^1)^2+ (\text{q}^2)^2 +(\text{q}^3)^2$. The follows can be checked:
\bea\la{A-58}
P^T P=P P^T=I,
\eea
where $I$ is $3\times 3$-unit matrix. So that Eq.(\ref{A-54}) can be rewritten as follows
\bea\nn
&&-\left(\nabla^T_xPP^T\mathfrak{B}^{(M)} PP^T\nabla_x\right)\phi_B=
-\left((\nabla^T_xP)(P^T\mathfrak{B}^{(M)} P)(P^T\nabla_x)\right)\phi_B\\
\la{A-59}&&\equiv -\left(\nabla^T_y\Lambda_B\nabla_y\right)\phi_B=-\eta\mathfrak{I} ={-4\pi e\over
\sqrt{-\det(B_{\mu\nu}^{(M)}(\mathcal{Q}))}}\delta^{(3)}(\mathbf{x}),
\eea
where
\bea\la{A-60}
&& y\equiv Px,~~{\rm or}~~y^i=P_{ij}x^j,~~x^i=P^T_{ij}y^j.\\
\la{A-61}&& \nabla_y=P^T\nabla_x,~~{\rm or}~~\pa/\pa y^i=P^T_{ij}\pa/\pa x^j,~~\pa/\pa x^i=P_{ij} \pa/\pa y^j.
\eea
Substituting Eqs.(\ref{A1-55}) (\ref{A-60}) into Eq.(\ref{A-59}) gives
\bea\nn
&&\hskip-0.2in\left[{\pa^2\over (\pa y^1/\sqrt{-\lambda_1})^2}+{\pa^2\over (\pa y^2/\sqrt{-\lambda_2})^2}+ {\pa^2\over (\pa y^3/\sqrt{-\lambda_3})^2}\right]\phi_B
=-\eta\delta(P^T_{1j}y^j)\delta(P^T_{2j}y^j)\delta(P^T_{3j}y^j)\\
\la{A-62}&&=-\eta {\delta(y^1)\delta(y^2)\delta(y^3)\over \left|\det(P^T_{ij})\right|}
={-4\pi e\over
\sqrt{-\det(B_{\mu\nu}^{(M)}(\mathcal{Q}))}}{\delta({y^1\over \sqrt{-\lambda_1}})\delta({y^2\over \sqrt{-\lambda_2}})\delta({y^3\over \sqrt{-\lambda_3}})\over \sqrt{-\lambda_1\lambda_2\lambda_3}},
\eea
where a calculation result $\left|\det(P^T_{ij})\right|=1$ has been used. Setting
\bea\la{A1-63}
\td{x}^i\equiv {y^i\over \sqrt{-\lambda^i}}~~{\rm with}~~i=1,\;2,\;3,
\eea
(\ref{A-62}) becomes
\bea\nn
&&\left[{\pa^2\over \pa (\td{x}^1)^2}+{\pa^2\over \pa (\td{x}^2)^2}+ {\pa^2\over \pa (\td{x}^3)^2}\right]\phi_B
\equiv \nabla_{\td{x}}^2\phi_B\\
\la{A1-64}&&={-4\pi e\over
\sqrt{-\det(B_{\mu\nu}^{(M)}(\mathcal{Q}))}}{\delta(\td{x}^1)\delta(\td{x}^2)\delta(\td{x}^3)\over \sqrt{-\lambda_1\lambda_2\lambda_3}}.
\eea
Then, we get the Coulomb potential:
\bea\la{A1-65}
\phi_B=\phi_B({\mathcal Q})={1\over
\sqrt{-\det(B_{\mu\nu}^{(M)}(\mathcal{Q}))}}{1\over \sqrt{-\lambda_1\lambda_2\lambda_3}}{e\over r_B},
\eea
where $r_B=\sqrt{(\td{x}^1)^2+(\td{x}^2)^2+(\td{x}^3)^2}$.


\end{enumerate}

\section{Fine Structure Constant Variation in $\{Q^0,\;Q^1,\;0,\;0\}$}
At this stage, in order to get the expression of fine-structure constant $\alpha(\mathcal{Q})\equiv \alpha_z(\Omega) $ at point $\mathcal{Q}\equiv\{Q^0,\;Q^1,\;Q^2,\;Q^3\}$, we should substitute Equations (\ref{6-57}), (\ref{2-5}), (\ref{AB4}), (\ref{A1-65}) and (\ref{1-1}) for $e_a^{~\mu},\;\omega_{~~\mu}^{ab},\;A^i,\;A^0\equiv\phi_B$ and $B_{\mu\nu}^{(M)}$ respectively, into the Dirac equation (\ref{Dirac0}) of hydrogen atom located in the local inertial coordinate system of the light-cone in FRW Universe with $\Lambda$ (see Fig.\ref{Fig1}). Such an $\alpha(\mathcal{Q})$ will characterize the temporal and spatial variations of fine-structure constant $\alpha$. When we assume $M=0$,  $\alpha(\mathcal{Q})|_{(M=0)}$ has been calculated in the Ref.\cite{Yan0,Yan01}. In the present paper we study the observation results of Keck and VLT reported by \cite{Webb1} recently  by means of $M=\{M^0,\;M^1,\;0,\;0\}$-model.

\begin{figure}[h]
\begin{center}
\includegraphics[scale=0.45]{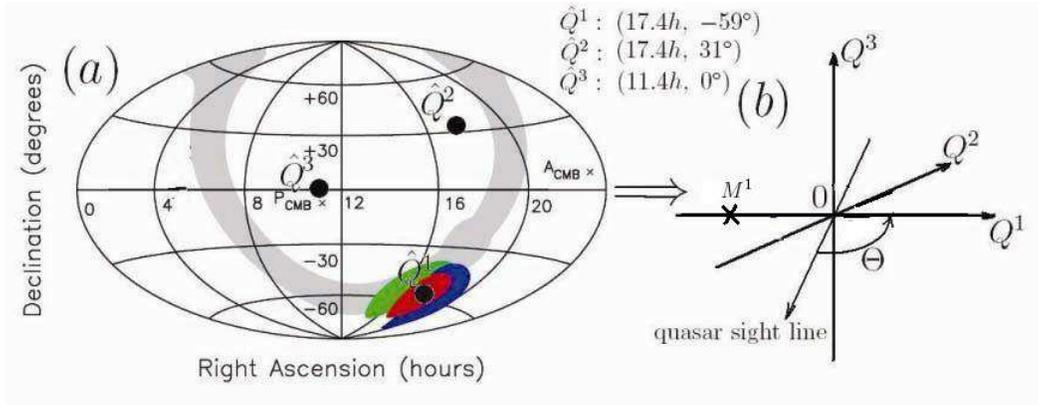}
\caption{\label{fig5} (color online) The 3-dimension spatial Cartesian coordinate frame
$\{Q^1,\;Q^2,\;Q^3\}$ on the equatorial coordinates. In left figure $(a)$, the background
 is all-sky plot in equatorial coordinates showing the independent Keck (green, leftmost)
and VLT (blue, rightmost) best-fit dipoles, and the combined samples (red, center), for
the dipole model $\Delta \alpha/\alpha=A\cos \Theta$ copied from \cite{Webb1}. The locations
of axis $\{\hat{Q}^1,\;\hat{Q}^2,\;\hat{Q}^3\}$ in this figure are marked by ``$\bullet$".
In \cite{Webb1} it has been measured that the best-fit dipole is at right ascension
$\varphi^{[RA]}=17.4\pm 0.9$ h, declination $\vartheta^{[D]}=-58\pm 9$ dec.
The cosmic microwave background dipole antipole are illustrated for comparison.
The directions of 3-dimension spatial Cartesian coordinate system $\{Q^1,\;Q^2,\;Q^3\}$
that we take are: $\hat{Q}^1(\varphi^{[RA]}_1,\vartheta^{[D]}_1)=\{17.4\;{\rm h},\;-59^\circ \} $,
 $\hat{Q}^2(\varphi^{[RA]}_2,\vartheta^{[D]}_2)=\{17.4\;{\rm h},\;31^\circ \} $,
$\hat{Q}^3(\varphi^{[RA]}_3,\vartheta^{[D]}_3)=\{11.4\;{\rm h},\;0^\circ \} $.
In right figure $(b)$, the 3-dimension spatial Cartesian coordinate system $\{Q^1,\;Q^2,\;Q^3\}$
 with a non-zero space component $M^1$ of Minkowski point is drawn.  $\Theta$ is angle between
a quasar sight line ($\{ \varphi^{[RA]}_q, \vartheta^{[D]}_q \})$ and axis $Q^1$. Formula for
computing it is $\cos \Theta=\cos[\vartheta^{[D]}_1] \cos[\vartheta^{[D]}_q]\cos[{\varphi^{[RA]}_1
-\varphi^{[RA]}_q\over 12}\pi]+ \sin[\vartheta^{[D]}_1] \sin[\vartheta^{[D]}_q]$.}
\end{center}
\end{figure}

To calculate $\alpha(\mathcal{Q})$ analytically, we build 3-dimension spatial Cartesian coordinate system $\{Q^1,\;Q^2,\;Q^3\}$ on the equatorial coordinate figure showing the Keck and VLT best-fit dipole structure of $\Delta \alpha/\alpha$ in Ref. \cite{Webb1}. Denoting $\hat{Q}^i\equiv Q^i/|\mathbf Q|$ and $Q^1$ as the best-fit dipole position, the directions of the three axis $\{\hat{Q}^1,\;\hat{Q}^2,\;\hat{Q}^3\}$ in this figure are $\hat{Q}^1(\varphi^{[RA]}_1,\vartheta^{[D]}_1)=\{17.4\;{\rm h},\;-59^\circ \} $, $\hat{Q}^2(\varphi^{[RA]}_2,\vartheta^{[D]}_2)=\{17.4\;{\rm h},\;31^\circ \} $, $\hat{Q}^3(\varphi^{[RA]}_3,\vartheta^{[D]}_3)=\{11.4\;{\rm h},\;0^\circ \} $ (see Figure \ref{fig5} and its caption). Angle ($\Theta$) between a quasar sight line ($\{ \varphi^{[RA]}_q, \vartheta^{[D]}_q \})$ and axis $Q^1$ is determined by following
\bea\la{A1-67+}
\cos \Theta=\cos[\vartheta^{[D]}_1] \cos[\vartheta^{[D]}_q]\cos[{\varphi^{[RA]}_1-\varphi^{[RA]}_q\over 12}\pi]+ \sin[\vartheta^{[D]}_1] \sin[\vartheta^{[D]}_q].
\eea
In this section we calculate $\alpha(\mathcal{Q})$ for $\Theta=\pi$ and $\Theta=0$ following the method in Ref. \cite{Yan0} and \cite{Yan01}.

\subsection{Formulation of Alpha-Variation for Case of $\Theta=\pi,$ and $0$}

When quasar sight line is anti-parallel (or parallel) to direction of $\hat{Q}^1(\varphi^{[RA]}_1,\vartheta^{[D]}_1)=\{17.4\;{\rm h},\;-59^\circ \} $, the angle $\Theta$ between them is equal to $\pi$ (or $0$), and the locations of distant atoms on the light-cone in the FRW Universe are at $\{Q^0,\;Q^1>0 \;(\rm{or}\;<0),\;Q^2=0,\;Q^3=0\}$ (see Figure \ref{Fig1}). Then we have $q^0=Q^0-M^0,\;q^1=Q^1-M^1,\;q^2=0, \;q^3=0$, and
\bea\la{AB1}
B_{\mu\nu}^{(M)}(Q)=\left(
\begin{array}{lccr}
{1\over \sigma^{(M)}}+{(Q^0-M^0)^2\over R^2(\sigma^{(M)})^2} & -{(Q^0-M^0)(Q^1-M^1)\over R^2(\sigma^{(M)})^2} &0 &0 \\
-{(Q^0-M^0)(Q^1-M^1)\over R^2(\sigma^{(M)})^2} & {-1\over \sigma^{(M)}}+{(Q^1-M^1)^2\over R^2(\sigma^{(M)})^2} & 0 & 0 \\
0 & 0 & {-1\over \sigma^{(M)}} & 0 \\
0 & 0 & 0 & {-1\over \sigma^{(M)}}
\end{array} \right),
\eea
\bea\la{AB2}
(B^{(M)})^{\mu\nu}(Q)=\left(
\begin{array}{lccr}
\sigma^{(M)}(1-{(Q^0-M^0)^2\over R^2}) & -{(Q^0-M^0)(Q^1-M^1)\sigma^{(M)}\over R^2} &0 &0 \\
-{(Q^0-M^0)(Q^1-M^1)\sigma^{(M)}\over R^2} & -\sigma^{(M)}(1+{(Q^1-M^1)^2\over R^2}) & 0 & 0 \\
0 & 0 & -\sigma^{(M)} & 0 \\
0 & 0 & 0 & -\sigma^{(M)}
\end{array} \right),
\eea
with
\bea\la{AB3}
\sigma^{(M)}=1-{(Q^0-M^0)^2-(Q^1-M^1)^2\over R^2}.
\eea
Here, for a known red-shift $z$, $Q^0\equiv c t(z)$ and $Q^1=\pm\sqrt{Q(z)^2-(Q^2)^2-(Q^3)^2}=\pm Q(z)$ are determined from Eqs.(\ref{newLa1}) and (\ref{5}) respectively (see also Fig.\ref{fig2} and Fig.\ref{fig3}).
Then Eqs.(\ref{A-56}) -(\ref{A-57}) become
\bea\la{A-72}
&&\lambda_1=-\sigma^{(M)}\left(1+\frac{(Q^1-M^1)^2 }{R^2}\right),~~\lambda_2=\lambda_3=-\sigma^{(M)},\\
\la{A-73} && P=\left(
\begin{array}{ccc}
 1 & 0 & 0 \\
 0 & 1 & 0 \\
 0 & 0 & 1
\end{array}
\right),
\eea
\noindent Substituting (\ref{6-57}) (\ref{2-5}) (\ref{AB4}) (\ref{A1-65}) and (\ref{1-1}) into (\ref{Dirac0}) gives
 dS-SR Dirac equation for the electron in the distant Hydrogen located at
 the light-cone of FRW Universe:
\begin{equation}\label{Dirac5}
\hbar c \beta \left[i\sqrt{\sigma^{(M)}}\gamma^\mu \mathcal{D}_\mu^L +i{\sigma^{(M)}-\sqrt{\sigma^{(M)}}\over (1-\sigma^{(M)})R^2} \eta_{ab}\delta^a_\lambda
(Q^\lambda-M^\lambda)\gamma^b (Q^\mu-M^\mu) \mathcal{D}_\mu^L-{m_e c\over \hbar}
\right]\psi=0,
\end{equation}
where factor $\hbar c \beta$ in the front of the equation is only
for convenience, $L^\mu\simeq Q^\mu$ has been used (see Fig.\ref{Fig1}), and
\bea\nn
\mathcal{D}_\mu^L&\equiv &{\pa\over \pa L^\mu}-{i\over
4}\omega^{ab}_{\;\;\mu}\sigma_{ab}-i{e\over c\hbar}B_{\mu\nu}^{(M)}A^\nu\\
\nn &=&
{\pa\over \pa L^\mu} - {i\over
4}{1\over R^2\left(1+\sqrt{\sigma^{(M)}}\right)\sqrt{\sigma^{(M)}}}(\delta^a_\mu\delta^b_\lambda-\delta^b_\mu\delta^a_\lambda )(Q^\lambda-M^\lambda)\sigma_{ab}\\
\la{+1} &&-i{e\over c\hbar}{\delta_{\mu 0}B_{00}^{(M)}(\mathcal{Q})+\delta_{\mu i}B_{i0}^{(M)}(\mathcal{Q})\over
\sqrt{-\det(B_{\mu\nu}^{(M)}(\mathcal{Q}))}}{1\over \sqrt{-\lambda_1\lambda_2\lambda_3}}{e\over r_B},
\eea
 We use the method suggested in \cite{Yan0} \cite{Yan01} and expand each terms of (\ref{Dirac5}) to the order as
follows:
\begin{enumerate}
\item Since observed distance hydrogen atom must be in the light cone and the location is $\{Q^0,\;Q^1,\;Q^2=0,\;Q^3=0\}$, then
$\eta_{ab}L^aL^b\simeq \eta_{ab}Q^aQ^b=(Q^0)^2-(Q^1)^2$, and the
first term of (\ref{Dirac5}) reads
\begin{eqnarray}
\label{first01} &&\hbar c \beta i\sqrt{\sigma^{(M)}}\gamma^\mu \mathcal{D}_\mu^L =
\sqrt{1-{(Q^0-M^0)^2-(Q^1-M^1)^2\over R^2}}
\hbar c \beta i \gamma^\mu \mathcal{D}^L_\mu \psi\\
\label{first02}&&{\rm with}~~\beta\gamma^\mu=\{\beta \gamma^0=\beta^2=1,\;\beta\gamma^i=\alpha^i\}\\
\label{first} &&\hbar c \beta i \gamma^\mu \mathcal{D}^L_\mu
\psi=(i\hbar {\pa\over \pa t_L} + i\hbar c \vec{\alpha}\cdot
\nabla + {\hbar c \beta \over 4}
\omega^{ab}_\mu\gamma^\mu\sigma_{ab} +e \beta\gamma^\mu
B_{\mu\nu}A^\nu ) \psi,
\end{eqnarray}
 In the follows, we use variables
$\{\td{x}^1, \td{x}^2, \td{x}^3\}$ (where since Eqs.(\ref{A-72})(\ref{A-73}), $\td{x}^i\equiv x^i/\sqrt{-\lambda^i}$, we have $\td{x}^1\equiv {1\over\sqrt{\sigma^{(M)}(1+(Q^1-M^1)^2/R^2)}}x^1$, $\td{x}^2\equiv {1\over\sqrt{\sigma^{(M)}}}x^2$ and $\td{x}^3\equiv {1\over\sqrt{\sigma^{(M)}}}x^3$ ) to replace $\{x^1, x^2, x^3\}$. Following
notations are introduced hereafter:
\begin{eqnarray}\label{nota1}
\mathbf{r}_B&=&\mathbf{i}\td{x}^1+ \mathbf{j}\td{x}^2+\mathbf{k}\td{x}^3,~~~~|\mathbf{r}_B|=r_B, \\
\label{nota2} \nabla_B &=& \mathbf{i}{\pa \over \pa\td{x}^1}+
\mathbf{j}{\pa \over \pa \td{x}^2}+\mathbf{k}{\pa\over \pa
\td{x}^3},~~~~~\td{x}^i\in\{\td{x}^1,\;\td{x}^2,\;\td{x}^3\}.
\end{eqnarray}
Then, noting ${\pa\over \pa x^1}={\pa \td{x}^1\over \pa
x^1}{\pa\over \pa \td{x}^1}={1\over \sqrt{\sigma^{(M)}(1+(Q^1-M^1)^2/R^2)}}{\pa\over \pa \td{x}^1}$, ${\pa\over \pa x^i}={\pa \td{x}^i\over \pa
x^i}{\pa\over \pa \td{x}^i}={1\over \sqrt{\sigma^{(M)}}}{\pa\over \pa \td{x}^i}$ with $i=\{2,\;3\}$, the eq.(\ref{first}) becomes {\footnotesize
\begin{eqnarray}
\nonumber &&\hbar c \beta i \gamma^\mu \mathcal{D}^L_\mu \psi
=\left(i\hbar {\pa\over \pa t_L} +i{\hbar c\over \sqrt{\sigma^{(M)}}} \vec{\alpha}\cdot \nabla_B+i{\hbar
c\over \sqrt{\sigma^{(M)}}}\left[{1\over \sqrt{1+(Q^1-M^1)^2/R^2}}-1\right]\alpha^1{\pa\over \pa\td{x}^1}\right.\\
\nn &&\hskip0.3in \left.+{\hbar c \beta \over 4}
\omega^{ab}_\mu\gamma^\mu\sigma_{ab}
 +eB_{00}^{(M)}\phi_B+e\alpha^1 B_{10}^{(M)} \phi_B \right) \psi\\
\nonumber &&\hskip0.1in=\left(i\hbar {\pa\over \pa t_L} +i{\hbar c\over \sqrt{\sigma^{(M)}}} \vec{\alpha}\cdot
\nabla_B+i{\hbar c\over \sqrt{\sigma^{(M)}}} \left[{1\over \sqrt{1+(Q^1-M^1)^2/R^2}}-1\right]\alpha^1{\pa\over \pa\td{x}^1}\right.\\
\label{first1} && \left.  +{\hbar c \beta \over 4}
\omega^{ab}_\mu\gamma^\mu\sigma_{ab}+\left[{1\over \sigma^{(M)}}+{(Q^0-M^0)^2\over R^2(\sigma^{(M)})^2} \right]e\phi_B-{(Q^1-M^1)(Q^0-M^0)\over R^2(\sigma^{(M)})^2}\alpha^1e\phi_B\right)  \psi.
\end{eqnarray} }

\item Estimation of the contributions of the fourth term in RSH of (\ref{first1})
( the spin-connection contributions) is as follows: From (\ref{2-5}), the ratio of
the fourth term to the first term of (\ref{first1}) is:
\begin{equation}\label{61}
\left|{ {\hbar c \beta \over 4}
\omega^{ab}_\mu\gamma^\mu\sigma_{ab}\psi \over i\hbar \pa_t\psi }
\right|\sim {\hbar c\over 4}{ct\over 2R^2}{1\over m_ec^2}={ct\over
8R^2}{\hbar\over m_ec}={1\over 8}{cta_c\over R^2}\sim 0,
\end{equation}
where $a_c= \hbar/(m_e c)\simeq 0.3\times 10^{-12}\rm{m}$ is the Compton
wave length of electron. $\mathcal{O}(cta_c/R^2)$-term is
neglectable. Therefore the 3-rd term in RSH of (\ref{first}) has no
contribution to our approximation calculations.

 \item Substituting  (\ref{61})
(\ref{first1}) (\ref{first02}) into (\ref{first01}) and noting
$L^a\simeq Q^a$ (see FIG. 1), we get the first term in LHS of
(\ref{Dirac5})
\begin{eqnarray}\nonumber
&&\hskip-0.2in \hbar c \beta i\sqrt{\sigma^{(M)}}\gamma^\mu \mathcal{D}_\mu^L \psi
=\sqrt{\sigma^{(M)}}\left(i\hbar {\pa\over \pa t_L} +i{\hbar c\over \sqrt{\sigma^{(M)}}} \vec{\alpha}\cdot
\nabla_B\right.\\
\nn && \left. +i{\hbar c\over \sqrt{\sigma^{(M)}}}
\left[{1\over \sqrt{1+(Q^1-M^1)^2/R^2}}-1\right]\alpha^1{\pa\over \pa\td{x}^1}\right.\\
\label{first-1} && \left.+\left[{1\over \sigma^{(M)}}+{(Q^0-M^0)^2\over R^2(\sigma^{(M)})^2}
 \right]e\phi_B-{(Q^1-M^1)(Q^0-M^0)\over R^2(\sigma^{(M)})^2}\alpha^1e\phi_B\right)  \psi.
\end{eqnarray}

\item The second term of (\ref{Dirac5}) is {\footnotesize
\begin{eqnarray}
 \nonumber &&\hskip-0.2in \hbar c \beta i {\sigma^{(M)}-\sqrt{\sigma^{(M)}} \over (1-\sigma^{(M)})R^2}
 \eta_{ab}(Q^a-M^a)\gamma^b (Q^\mu-M^\mu) \mathcal{D}_\mu^L\psi\\
\nn &&=\hbar c \beta i{\sigma^{(M)}-\sqrt{\sigma^{(M)}} \over (1-\sigma^{(M)})R^2} \left[\gamma^0
(Q^0-M^0)-\vec{\gamma}\cdot (\vec{Q}-\vec{M})\right] \left[(Q^0-M^0)\mathcal{D}_0
+(Q^i-M^i)\mathcal{D}_i \right]\psi\\
\nonumber  && = \hbar c  i{\sigma^{(M)}-\sqrt{\sigma^{(M)}} \over (1-\sigma^{(M)})R^2}
\left[(Q^0-M^0)-\alpha^1(Q^1-M^1)\right]\\
\nn && \hskip0.1in \times\left[ (Q^0-M^0)\left(\pa_0^L-{\gamma^0\gamma^1 (Q^1-M^1)\over
2R^2(1+\sqrt{\sigma^{(M)}})\sqrt{\sigma^{(M)}}}-{ie\over c\hbar}({1\over\sigma^{(M)}}+{(Q^0-M^0)^2\over R^2(\sigma^{(M)})^2})\phi_B\right)\right.\\
\nn &&\hskip0.2in \left. +(Q^1-M^1)\left(\pa_1^L+{\gamma^0\gamma^1 (Q^0-M^0)\over
2R^2(1+\sqrt{\sigma^{(M)}})\sqrt{\sigma^{(M)}}}+{ie\over c\hbar}{(Q^0-M^0)(Q^1-M^1)\over R^2(\sigma^{(M)})^2}\phi_B\right)\right]\psi \\
\nonumber &&={\sigma^{(M)}-\sqrt{\sigma^{(M)}} \over (1-\sigma^{(M)})R^2}
\left[(Q^0-M^0)-\alpha^1(Q^1-M^1)\right]\\
\la{second} && \hskip0.2in \times \left\{i\hbar c\left[(Q^0-M^0){\pa\over \pa L^0}+(Q^1-M^1){\pa\over \pa L^1}\right]
+{Q^0-M^0\over (\sigma^{(M)})^2}e\phi_B\right\}\psi,
\end{eqnarray} }
where $Q^2=Q^3=M^2=M^3=0$, $\gamma^0=\beta,\;\beta\gamma^1=\alpha^1$ and Eqs.(\ref{+1}) (\ref{AB3})
 were used.
Noting $x^\mu\equiv L^\mu-Q^\mu$, ${\pa\over \pa L^0}={\pa\over \pa x^0}={\pa\over c\pa t_L}$,
${\pa\over \pa L^1}={\pa\over \pa x^1}={\pa \td{x}^1\over \pa
x^1}{\pa\over \pa \td{x}^1}={1\over \sqrt{\sigma^{(M)}(1+(Q^1-M^1)^2/R^2)}}{\pa\over \pa \td{x}^1}$,
 (\ref{second}) becomes
\begin{eqnarray}
 \nonumber &&\hskip-0.2in \hbar c \beta i {\sigma^{(M)}-\sqrt{\sigma^{(M)}} \over (1-\sigma^{(M)})R^2}
 \eta_{ab}\delta^a_\lambda(Q^\lambda-M^\lambda)\gamma^b (Q^\mu-M^\mu) \mathcal{D}_\mu^L\psi\\
\nonumber &&={\sigma^{(M)}-\sqrt{\sigma^{(M)}} \over (1-\sigma^{(M)})R^2}
\left[(Q^0-M^0)-\alpha^1(Q^1-M^1)\right]\\
\nn && \hskip0.2in \times \left\{\left[(Q^0-M^0)i\hbar{\pa\over \pa t_L}+(Q^1-M^1)
{i\hbar c\over \sqrt{\sigma^{(M)}(1+(Q^1-M^1)^2/R^2)}}{\pa\over \pa \td{x}^1}\right]\right.\\
\la{second1}&&\hskip0.6in +\left.{Q^0-M^0\over (\sigma^{(M)})^2}e\phi_B\right\}\psi.
\end{eqnarray}

\item Therefore, substituting (\ref{first-1}) (\ref{second1}) into (\ref{Dirac5}), we
have
\begin{eqnarray}\nonumber
&& i\hbar\left(\sqrt{\sigma^{(M)}}+{\sigma^{(M)}-\sqrt{\sigma^{(M)}}\over (1-\sigma^{(M)})R^2}[(Q^0-M^0)^2-(Q^0-M^0)(Q^1-M^1)\alpha^1]\right){\pa\over \pa t_L}\psi \\
\nn &&\hskip-0.4in =\left\{-i\hbar c\vec{\alpha}\cdot\nabla_B-\left[{1\over\sqrt{\sigma^{(M)}}}+{(Q^0-M^0)^2\over R^2\sigma^{(M)}\sqrt{\sigma^{(M)}}}+{\sigma^{(M)}-\sqrt{\sigma^{(M)}}\over (1-\sigma^{(M)})R^2}{(Q^0-M^0)^2\over(\sigma^{(M)})^2}\right]e\phi_B + m_e c^2\beta \right\}\psi\\
 \nn &&+\left\{-i\hbar c\left({1\over \sqrt{1+(Q^1-M^1)^2/R^2}}-1\right)\alpha^1{\pa\over \pa\td{x}^1}+{(Q^1-M^1)(Q^0-M^0)\over R^2\sigma^{(M)}\sqrt{\sigma^{(M)}}}\alpha^1e\phi_B\right.\\
\label{Dirac6} && \hskip-0.2in \left. -{\sigma^{(M)}-\sqrt{\sigma^{(M)}}\over (1-\sqrt{\sigma^{(M)}})R^2} \left[{i\hbar c(Q^1Q^0-(Q^1-M^1)^2\alpha^1)\over \sqrt{\sigma^{(M)}(1+(Q^1-M^1)^2/R^2)}} {\pa\over\pa \td{x}^1} \right]\right\}\psi.
\end{eqnarray}
In order to discuss the spectra of Hydrogen atom in the dS-SR Dirac equation, we need to find its solutions with certain energy $E$ for electron in the atom. From Eq.(68) in Ref.\cite{Ours}, we have
\bea\nn
&& p^0\equiv{E\over c}=i\hbar\left[\left(\eta^{0\nu}-{(Q^0-M^0)(Q^\nu-M^\nu)\over R^2}\right)\pa_\nu+{5(Q^0-M^0)\over 2R^2}\right],\\
\nn &&E=i\hbar\left[\pa_{t_L}-{(Q^0-M^0)^2\over R^2}\pa_{t_L}-{c(Q^0-M^0)(Q^1-M^1)\over R^2}\pa_{L^1}+{5c(Q^0-M^0)\over 2R^2}\right],\\
\nn &&E\psi\simeq i\hbar\left(1-{(Q^0-M^0)^2\over R^2}\right)\pa_{t_L}\psi
-i\hbar{c(Q^0-M^0)(Q^1-M^1)\over R^2\sqrt{\sigma^{(M)}(1+{(Q^1-M^1)^2/R^2}}}{\pa\over \pa\td{x}^1}\psi,\\
\la{+2}
\eea
where a estimation for the ratio of the 4-th term to the 2-nd of
$E\psi$ were used:
$${|i\hbar {5c^2t\over 2R^2}\psi|\over |{-c^2t^2\over R^2}i\hbar
\pa_{t_L}\psi|}\sim {|i\hbar {5c^2t\over 2R^2}|\over |{-2c^2t^2\over
R^2}E|}\sim {5\hbar \over 2t m_e c^2}\equiv {5\over 2}{a_c\over ct}
$$  where $a_c\equiv \hbar/(m_ec)\simeq 0.3\times 10^{-12}$m is the Compton wave length of electron and $ct$ is about the
distance between earth and a distant atom near quasar. Obviously,
$a_c/(ct)$ is ignorable. For instance, to a atom with $ct\sim
10^9$ly, $a_c/(ct)\sim 10^{-38}<< (ct)^2/R^2\sim 10^{-5}$. Hence the
4-th term of $E\psi$ were ignored. Eq.(\ref{+2}) means
\bea\nn
i\hbar{\pa\over \pa t_L}\psi&=&{E\over 1-(Q^0-M^0)^2/R^2}\psi\\
\la{+3} &&+{i\hbar c(Q^1-M^1)(Q^0-M^0)\over R^2(1-(Q^0-M^0)^2/R^2)}{1\over\sqrt{\sigma^{(M)}(1+(Q^1-M^1)^2/R^2)}}{\pa\over\pa \td{x}^1}\psi.
\eea
Then substituting (\ref{+3}) into (\ref{Dirac6}) gives
{\footnotesize
\begin{eqnarray}\label{rrDirac6}
&&E\psi=H_0\psi+H'\psi\\
\nn{\rm with}&& \hskip-0.2in H_0=\left(1-{(Q^0-M^0)^2\over R^2}\right)\left[\sqrt{\sigma^{(M)}}+{(\sigma^{(M)}-\sqrt{\sigma^{(M)}})(Q^0-M^0)^2\over (1-\sigma^{(M)})R^2}\right]^{-1}\\
\nn &&\times \left\{-i\hbar c\vec{\alpha}\cdot\nabla_B-\left[{1\over\sqrt{\sigma^{(M)}}}+{(Q^0-M^0)^2\over R^2\sigma^{(M)}}\left({1\over 1+\sqrt{\sigma^{(M)}}}\right)\right]e\phi_B + m_e c^2\beta \right\}\\
\la{H0} && \equiv -i\hbar_z(\Omega)c \vec{\alpha}\cdot\nabla_B-{e_z(\Omega)^2\over r_B}+m_{e,\;z}(\Omega) c^2\beta,\\
\nn && \hskip-0.4in H'=\left(1-{(Q^0-M^0)^2\over R^2}\right)\left[\sqrt{\sigma^{(M)}}+{(\sigma^{(M)}-\sqrt{\sigma^{(M)}})(Q^0-M^0)^2\over (1-\sigma^{(M)})R^2}\right]^{-1}\\
\nn && \times\left\{-i\hbar c\left({1\over \sqrt{1+(Q^1-M^1)^2/R^2}}-1\right)\alpha^1{\pa\over \pa\td{x}^1}\right.+{(Q^1-M^1)(Q^0-M^0)\over R^2\sigma^{(M)}\sqrt{\sigma^{(M)}}}\alpha^1e\phi_B\\
\nn && \left. \hskip-0.5in -{\sigma^{(M)}-\sqrt{\sigma^{(M)}}\over (1-\sigma^{(M)})R^2} \left(i\hbar c((Q^1-M^1)(Q^0-M^0)-(Q^1-M^1)^2\alpha^1){\pa\over\pa \td{x}^1} -{(Q^1-M^1)(Q^0-M^0)^2\over (\sigma^{(M)})^2} \alpha^1e\phi_B\right)\right.\\
\nn && \left. +{(\sigma^{(M)}-\sqrt{\sigma^{(M)}})(Q^0-M^0)(Q^1-M^1)\alpha^1\over (1-(Q^0-M^0)^2/R^2)(1-\sigma^{(M)})R^2}\left(E+i\hbar c{(Q^0-M^0)(Q^1-M^1)\over R^2\sqrt{1+(Q^1-M^1)^2/R^2}}{\pa\over\pa \td{x}^1}\right)\right\}\\
\nn && -{i\hbar c (Q^1-M^1)(Q^0-M^0)\over R^2\sqrt{\sigma^{(M)}(1+(Q^1-M^1)^2/R^2)}}{\pa\over \pa \td{x}^1}\\
\la{H1}&\equiv &\sum_{i=1}^3 C_i \hat{O}^i,
\end{eqnarray} }
where
\bea\nn
&& \hat{Q}^1=\alpha^1,~~~C_1\propto \mathcal{O}(1/R^2);~~~~~\hat{Q}^2={\pa\over \pa\td{x}^1},~~~C_2\propto \mathcal{O}(1/R^2);\\
\la{100}&& \hat{Q}^3=\alpha^1{\pa\over \pa\td{x}^1},~~~C_3\propto \mathcal{O}(1/R^4),
\eea
and so that
\bea\la{101}
H'\propto \mathcal{O}(1/R^2)<< H_0 \propto \mathcal{O}(1).
\eea
We have mentioned in the Introduction section that the operator-structure of $H_0$ of (\ref{H0}) makes the corresponding eigen-equation $E_0\psi=H_0\psi$ to be integrable. Hence Eq.(\ref{101}) means that $H_0$ (Eq.(\ref{H0})) and $H'$ (Eq.(\ref{H1})) can be legally treated as unperturbation Hamiltonian and perturbation Hamiltonian respectively in QM-problem with $H=H_0+H'$.

\item From Eq.(\ref{H0}), we have
\bea\la{102}
&& \hbar_z(\Omega)c=\hbar c\left(1-{(Q^0-M^0)^2\over R^2}\right)\hskip-0.1in \left[\sqrt{\sigma^{(M)}}+{(\sigma^{(M)}-\sqrt{\sigma^{(M)}})(Q^0-M^0)^2\over (1-\sigma^{(M)})R^2}\right]^{-1}\hskip-0.15in,\\
\nn && e^2_z(\Omega)=e^2\left(1-{(Q^0-M^0)^2\over R^2}\right)\left[\sqrt{\sigma^{(M)}}+{(\sigma^{(M)}-\sqrt{\sigma^{(M)}})(Q^0-M^0)^2\over (1-\sigma^{(M)})R^2}\right]^{-1}\\
\nn &&\times \left\{\left[{1\over\sqrt{\sigma^{(M)}}}+{(Q^0-M^0)^2\over R^2\sigma^{(M)}}\left({1\over 1+\sqrt{\sigma^{(M)}}}\right)\right]{1\over
\sqrt{-\det(B_{\mu\nu}^{(M)}(\mathcal{Q}))}}{1\over \sqrt{-\lambda_1\lambda_2\lambda_3}}  \right\},\\
\la{103}
\eea
where $\phi_B$-expression (\ref{A1-65}) were used, and hence
\bea\la{104}
\alpha_z(\Omega) =\alpha {{1\over\sqrt{\sigma^{(M)}}}+{(Q^0-M^0)^2\over R^2\sigma^{(M)}}\left({1\over 1+\sqrt{\sigma^{(M)}}}\right)\over
\sqrt{-\det(B_{\mu\nu}^{(M)}(\mathcal{Q}))} \sqrt{-\lambda_1\lambda_2\lambda_3}},~~{\rm with}~~ \alpha={e^2\over \hbar c}.
\eea
For case of $\Theta=\pi,({\rm or}\;0)$, $B_{\mu\nu}^{(M)}(\mathcal{Q})$ and $\lambda_1,\; \lambda_2,\;\lambda_3$ have been given in Eqs.(\ref{AB1}) and (\ref{A-72}) respectively, and hence we have
\bea\nn
&&\sqrt{-\det(B_{\mu\nu}^{(M)}(\mathcal{Q}))}\\
\nn &=&{1\over \sigma^{(M)}}\sqrt{ \left({1\over \sigma^{(M)}}-{(Q^1-M^1)^2\over R^2(\sigma^{(M)})^2}\right) \left({1\over\sigma^{(M)}}+{(Q^0-M^0)^2\over R^2(\sigma^{(M)})^2}\right)+{(Q^0-M^0)^2(Q^1-M^1)^2\over R^4(\sigma^{(M)})^4}},\\
\la{105}\\
\la{106}&& \sqrt{-\lambda_1\lambda_2\lambda_3}=(\sigma^{(M)})^{3/2}\sqrt{1+{(Q^1-M^1)^2
\over R^2}}.
\eea
Substituting (\ref{105}) (\ref{106}) into (\ref{104}) gives
\bea\nn \alpha_z(\Omega)\hskip-0.1in & =&\hskip-0.1in\alpha
{ {1\over\sigma^{(M)}}+{(Q^0-M^0)^2\over R^2\sigma^{(M)}\sqrt{\sigma^{(M)}}}\left({1\over 1+\sqrt{\sigma^{(M)}}}\right) \over \sqrt{\left[ \left({1\over \sigma^{(M)}}-{(Q^1-M^1)^2\over R^2(\sigma^{(M)})^2}\right) \left({1\over\sigma^{(M)}}+{(Q^0-M^0)^2\over R^2(\sigma^{(M)})^2}\right)+{(Q^0-M^0)^2(Q^1-M^1)^2\over R^4(\sigma^{(M)})^4}\right] \left(1+{(Q^1-M^1)^2
\over R^2}\right)}}.\\
\la{107}
\eea
When $z=0$ (or $Q^0\rightarrow 0$, and $Q^1\rightarrow 0$), $\alpha_z(\Omega)$ should be normalized to be $\alpha_z(\Omega)|_{z=0}=\alpha_0$ which is the $\alpha$-value measured in the Earth laboratories. So from (\ref{107}) we have
\bea\nn
\alpha_0&=&\alpha
{ {1\over\sigma^{(M)}_0}+{(M^0)^2\over R^2\sigma^{(M)}_0\sqrt{\sigma^{(M)}_0}}\left({1\over 1+\sqrt{\sigma^{(M)}_0}}\right) \over \sqrt{\left[ \left({1\over \sigma^{(M)}_0}-{(M^1)^2\over R^2(\sigma^{(M)}_0)^2}\right) \left({1\over\sigma^{(M)}_0}+{(M^0)^2\over R^2(\sigma^{(M)}_0)^2}\right)+{(M^0)^2(M^1)^2\over R^4(\sigma^{(M)}_0)^4}\right] \left(1+{(M^1)^2
\over R^2}\right)}}\\
\la{108}&\equiv &\alpha N_0
\eea
where
\bea\la{109}
&&\hskip-0.4in N_0={ {1\over\sigma^{(M)}_0}+{(M^0)^2\over R^2\sigma^{(M)}_0\sqrt{\sigma^{(M)}_0}}\left({1\over 1+\sqrt{\sigma^{(M)}_0}}\right) \over \sqrt{\left[ \left({1\over \sigma^{(M)}_0}-{(M^1)^2\over R^2(\sigma^{(M)}_0)^2}\right) \left({1\over\sigma^{(M)}_0}+{(M^0)^2\over R^2(\sigma^{(M)}_0)^2}\right)+{(M^0)^2(M^1)^2\over R^4(\sigma^{(M)}_0)^4}\right] \left(1+{(M^1)^2
\over R^2}\right)}}\\
\la{110}&&\hskip-0.4in \sigma^{(M)}_0\equiv \left.\sigma^{(M)}\right|_{Q^0=Q^1=0}=1-{(M^0)^2-(M^1)^2\over R^2}.
\eea
Therefore, from (\ref{107}) and (\ref{108}), we obtain
\bea\nn
&&\hskip-0.4in {\Delta \alpha\over \alpha_0}\equiv {\alpha_z(\Omega)-\alpha_0\over \alpha_0}\\
\nn &&\hskip-0.4in ={ {1\over\sigma^{(M)}}+{(Q^0-M^0)^2\over R^2\sigma^{(M)}\sqrt{\sigma^{(M)}}}\left({1\over 1+\sqrt{\sigma^{(M)}}}\right) \over N_0\sqrt{\left[ \left({1\over \sigma^{(M)}}-{(Q^1-M^1)^2\over R^2(\sigma^{(M)})^2}\right) \left({1\over\sigma^{(M)}}+{(Q^0-M^0)^2\over R^2(\sigma^{(M)})^2}\right)+{(Q^0-M^0)^2(Q^1-M^1)^2\over R^4(\sigma^{(M)})^4}\right] \left(1+{(Q^1-M^1)^2
\over R^2}\right)}}-1,\\
\la{111}
\eea
where $Q^0(z)\equiv c t(z)$ and $Q^1(z)=\pm\left.\sqrt{Q(z)^2-(Q^2)^2-(Q^3)^2}\right|_{(Q^2=Q^3=0)}=\pm Q(z)$ (i.e., $|Q^1(z)|=Q(z)$) have been given in Eqs.(\ref{newLa1}) and (\ref{5}) respectively (see also Fig.\ref{fig2} and Fig.\ref{fig3}).

\end{enumerate}

\subsection{Comparisons with Observations of Alpha-varying for Cases of $\Theta=0,\;\pi$}
Equation (\ref{111}) is the prediction of $\alpha$-varying derived from dS-SR Dirac equation of distant hydrogen atom located in $Q^1$-axis. When the location $Q^1>0$, the direction of the corresponding quasar sight line is opposite to the direction of $Q^1$-axis (see Fig.\ref{fig5}). For this case the corresponding angle ( $\Theta$ ) between the quasar sight line and the $Q^1$-axis is equal to $\pi$ (i.e., $\Theta=\pi$ for this case). Oppositely, when the location $Q^1<0$, we have $\Theta=0$, i.e., the direction of the quasar sight line is same to the direction of $Q^1$-axis. For both of cases, we can generally write $Q^1$ in Eq.(\ref{111}) as
\bea\la{112}
Q^1=-|Q^1|\cos \Theta=-Q(z)\cos\Theta.
\eea
Looking back cosmic time variable is $Q^0\equiv ct<0$, and the coordinates of Minkowski point of Beltrami metric $B^{(M)}_{\mu\nu}$ are $M^0<0$, $M^1>0$, and $M^2=M^3=0$. Substituting (\ref{111}) into (\ref{111}) gives
{\footnotesize
\bea\nn
{\Delta \alpha\over \alpha_0}
 &\hskip-0.12in=&\hskip-0.12in{\left[ {1\over\sigma^{(M)}}+{(Q^0(z)-M^0)^2\over R^2\sigma^{(M)}\sqrt{\sigma^{(M)}}}\left({1\over 1+\sqrt{\sigma^{(M)}}}\right)\right]/N_0 \over \sqrt{\hskip-0.03in\left[ \hskip-0.03in\left({1\over \sigma^{(M)}}-{(-Q(z)\cos\Theta-M^1)^2\over R^2(\sigma^{(M)})^2}\right) \hskip-0.07in\left({1\over\sigma^{(M)}}+{(Q^0(z)-M^0)^2\over R^2(\sigma^{(M)})^2}\right)\hskip-0.05in +\hskip-0.05in {(Q^0(z)-M^0)^2(-Q(z)\cos\Theta-M^1)^2\over R^4(\sigma^{(M)})^4}\right] \hskip-0.07in\left(1+{(-Q(z)\cos\Theta-M^1)^2
\over R^2}\right)}}\\
\la{113} &&-1,
\eea
where $N_0$ is the same as (\ref{109}) and
\bea\la{114}
\sigma^{(M)}=1-{(Q^0(z)-M^0)^2-(-\cos\Theta Q(z)-M^1)^2\over R^2}.
\eea }
The equation (\ref{113}) is the theoretic prediction of dS-SR, whose variables are $z$ (red shift) and $\Theta(=0,\;{\rm or}\;\pi)$, and the adjustable parameters are $R,\;M^0,\;M^1$. The discussions on it are follows:
\begin{enumerate}
\item Since $|R|$ is the maximal length scale parameter in the theory (say $|R|\sim 10^{12}$lyr \cite{Yan+}), we could deduce the ${\Delta\alpha/\alpha_0}$'s Taylor-power series of $1/R^2$ from (\ref{113}) (for practical calculations, {\it ``Mathematica"} is useful):
\bea\nn
&&{\Delta\alpha\over \alpha_0}\simeq\frac{1}{8R^4} \left[(M^1)^2 Q^0(z)(-2 M^0+Q^0(z))+2 M^1 (M^0-Q^0(z))^2 Q(z) \cos\Theta \right.\\
\la{115}&&\hskip0.4in \left.+(M^0-Q^0(z))^2 (Q(z))^2 \cos^2\Theta\right] +\mathcal{O}(1/R^6),
\eea
where the leading term $\propto \mathcal{O}(1/R^4)$ is dominating in the expansion of Eq.(\ref{113}). Suppose the parameters and the variables are chosen such that
\bea\la{116}
|M^0|>>|M^1|,~~~Q^0(z)\sim Q(z)\sim \epsilon,
\eea
we have
\bea\la{117}
{\Delta\alpha\over \alpha_0}\sim \frac{1}{8R^4} 2 \cos\Theta\; M^1 (M^0)^2\epsilon,
\eea
where the lesser terms $\propto \mathcal{O}((M^1)^2M^0\epsilon/R^4)$ and $\propto\mathcal{O}(\epsilon^2/R^4)$ have been ignored, and $\Theta$ is only to be $0$ or $\pi$ (and noting $\cos(0)=1,\;\cos(\pi)=-1$). Thus, when $\Theta=0$, we have $\Delta\alpha/\alpha_0>0$, and when $\Theta=\pi$, oppositely, we have $\Delta\alpha/\alpha_0<0$. This is interesting since Eq.(\ref{117}) indicates that the scenario reported by \cite{Webb1} could be interpreted by (\ref{113}) with a particular parameter setting in proper region of variables (\ref{116}) in the theory.

\item In this scenario we need to determine the parameters $R,\;M^0,\;M^1$ by comparing the theoretical predictions with the observation data. Keck+VLT data have shown the relations between $\Delta\alpha/\alpha_0$ and $r\equiv ct(z)$ along $\hat{Q}^1$-axis \cite{Webb1}.
Let's use Eqs.(\ref{113}) and (\ref{114}) with $\Theta=\pi$ and $0$ to fit the figure 3 in \cite{Webb1} which is based on the combined Keck and VLT data and $t(z)$ expression (\ref{newLa1}).  The best fitting gives
\bea\nn
&& R=500\;{\rm Glyr}=0.5\times 10^{12}\;{\rm Lyr},\\
\la{120}&& M^0\simeq -100\;{\rm Glyr}=-1.0\times 10^{11}\;{\rm Lyr},\\
\nn && M^1\simeq -22\;{\rm Glyr}=-2.2\times 10^{10}\;{\rm Lyr},
\eea
which are consistent with requirement of (\ref{116}).
The fitted curve of (\ref{113}) with (\ref{114}) is shown in Fig \ref{fitting}.
\begin{figure}[h]
\begin{center}
\includegraphics[scale=0.5]{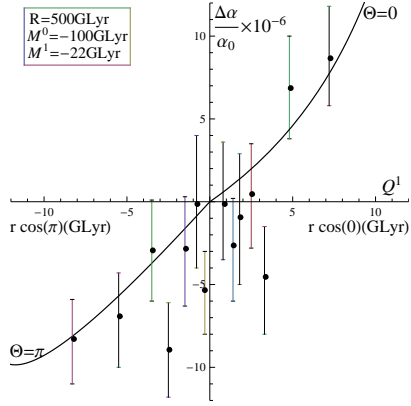}
\caption{\label{fitting} \footnotesize Determination of parameters $R,\;M^0,\;M^1$ via fitting Keck+VLT's $\alpha$-varying data reported by \cite{Webb1}. ${\Delta\alpha/\alpha_0}$ vs $A\;r(z)cos\Theta$ with $\Theta=\{0,\;{\rm and}\;\pi\}$ shows an apparent gradient in $\alpha$ along the best-fit dipole. The best-fit direction is at $\hat{Q}^1(\varphi^{[RA]}_1,\vartheta^{[D]}_1)=\{17.4\;{\rm h},\;-59^\circ \}$ (see fig.\ref{fig5}). The data reported by \cite{Webb1} are shown with error bars.
 A spatial gradient is
statistically preferred over a monopole-only model at the
4.2$\sigma$ level. A cosmology with parameters $(H_0,\;\Omega_M,\;\Omega_\Lambda)$ were given in (\ref{newLa4}). The fitted  model's parameters are $R\simeq 500 {\rm GLyr},\;M^0\simeq-100 {\rm GLyr},\; M^1\simeq -22 {\rm GLyr} $. The resulting curve of $\Delta\alpha/\alpha_0(r(z))$ of Eq.(\ref{113}) is shown.  }
\end{center}
\end{figure}

\item After determination of $R,\;M^0,\;  M^1$ of Eq.(\ref{120}), the metric $B^{(M)}_{\mu\nu}(Q)$ of local inertial coordinate system and then $\Delta\alpha/\alpha_0(z)$ along $Q^1$-axis are fully known. In figure \ref{fig7}, the curves of $\Delta\alpha/\alpha_0(z)$ are plotted. The Keck's data in 2004 reported by \cite{Webb04} \cite{Dent} are illustrated for comparison. Since the 2004-data were obtained by observations in all directions in Keck at that time, the deviations between the data and the prediction curves of $\Delta\alpha/\alpha_0(z)$ are understandable. The point here is that the curves remarkably show a nontrivial scenario described and reported by Ref.\cite{Webb1}. That is, in one direction in the sky $\alpha$ was
smaller at the time of absorption, while in the opposite direction it was larger. More explicitly, we illustrate 3 pairs of $\Delta\alpha/\alpha_0(z)$-predictions in table 1 as examples. In the table, $\Theta=0$ (or $\pi$) means the quasar sight line is parallel (or anti-parallel) to direction of $\hat{Q}^1(\varphi^{[RA]}_1,\vartheta^{[D]}_1)=\{17.4\;{\rm h},\;-59^\circ \} $. For each $z$, there are two values of $(\Delta\alpha/\alpha_0)_{\rm th}$ with opposite signs, which just matches the expectations of observations. Such a theoretical picture is subtle.
  In addition, it were also reported as a dipole form in \cite{Webb1}
\bea\la{121}
{\Delta\alpha\over\alpha_0}\simeq \bar{A}_{\rm obs}\cos\Theta, ~~~~{\rm with}~~\bar{A}_{\rm obs}=(1.02 \pm 0.21)\times 10^{-5},
\eea
where $\bar{A}_{obs}$ means the observation value of amplitude $\bar{A}$. Theoretically, figure \ref{fig7} indicates that when $z\sim 2$ to 4, $\bar{A}_{\rm th}\simeq 1. \times 10^{-5}$ which coincides with $\bar{A}_{\rm obs}$.
\begin{figure}[h]
\begin{center}
\includegraphics[scale=0.7]{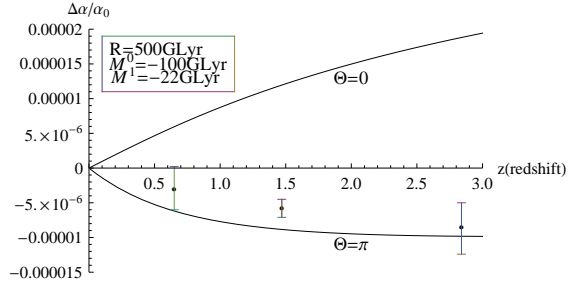}
\caption{\label{fig7}  $\alpha$-varying in the region of $0<z<3$. $\Theta$ is angle between quasar sight line and axis $Q^1$. $\Theta=0$ (or $\pi$) means the quasar sight line is parallel (or anti-parallel) to direction of $\hat{Q}^1(\varphi^{[RA]}_1,\vartheta^{[D]}_1)=\{17.4\;{\rm h},\;-59^\circ \} $. When $z$ fixes, there are two values of $\Delta\alpha/\alpha_0(z)$ with opposite signs. $\Delta\alpha/\alpha_0(z)$ is given by Eq.(\ref{113}) with parameters Eq.(\ref{120}). Three Keck's data in 2004 reported and discussed by \cite{Webb04} \cite{Dent} are shown for comparison.}
\end{center}
\end{figure}

\begin{table}
\caption{ Examples of predictions of $\Delta \alpha/\alpha_0$: $\Theta=\{0,\;\pi\}$ is angle between quasar sight line and axis $Q^1$. For each redshift $z$, there is a pair of $(\Delta \alpha/\alpha_0)_{\rm th}$-predictions from Eq.(\ref{113}) with parameters in Eq.(\ref{120}).}
\tabcolsep 0.1in
\begin{tabular}{c| c |c |c| c| c| c } \hline \hline
  $ z$ & \multicolumn{2}{|c|}{0.65} & \multicolumn{2}{|c|}{1.47} &\multicolumn{2}{|c}{2.84}  \\ \hline
 $\Theta$ & 0 & $\pi$ &  0 & $\pi$ & 0&  $\pi$  \\
 $(\Delta\alpha/\alpha_0)_{th}$ &{\footnotesize $0.60\times 10^{-5}$}&{\footnotesize $-0.62\times 10^{-5}$}& {\footnotesize $1.20\times10^{-5}$} &{\footnotesize$-0.88\times 10^{-5}$ }&{\footnotesize $1.88\times10^{-5}$ }&{\footnotesize $-0.98\times 10^-6$}
\\\hline \hline
\end{tabular}
\end{table}

\item We further plot the curves of $\Delta\alpha/\alpha_0$ of (\ref{113}) with (\ref{120}) in more wide $z$-region including radiation
epoch of the Universe in Figure \ref{fit2} (that epoch roughly corresponds to $z \geq 3 \times 10^3$).
For $\Theta=0$, the radiation epoch limit of $\Delta\alpha/\alpha$ is $4.7\times 10^{-5}$, while for $\Theta=\pi$, this limit is about $-5\times 10^{-6}$. Therefore, we find that in that epoch the dipole form (\ref{121}) is no longer true.

\begin{figure}[h]
\begin{center}
\includegraphics[scale=1]{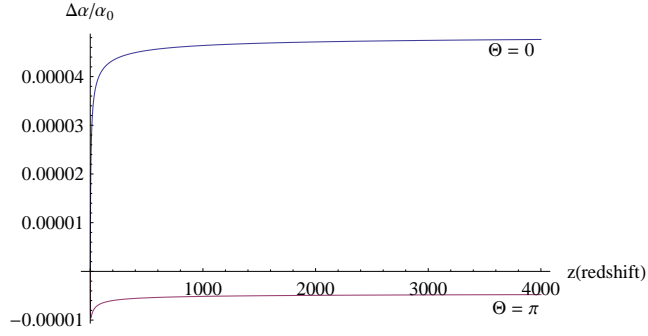}
\caption{\label{fit2} \small $\alpha$-varying in the region of $0<z<4000$. }
\end{center}
\end{figure}

\end{enumerate}

\section{$\alpha$-Varying in Whole Sky}
In the last section, the $\alpha$-varying for $\Theta=\{0,\;\pi\}$ (or for case that both distant atom and quasar lie at $Q^1$-axis) was studied and the model's parameters $R,\;M^0,\;M^1$ have been determined. Now we discuss general case of $\Theta\in \{0,\;\pi\}$, i.e., the case of $Q^1\neq 0,$ $Q^2\neq 0,$ and $Q^3\neq0$ (see Fig. \ref{fig5}). The corresponding M-Beltrami metrics (\ref{01-4}) reads
\bea\nn
\{B^{(M)}_{\mu\nu}\}=\left(
\begin{array}{cccc}
 \frac{(Q^0-M^0)^2}{R^2 (\sigma^{(M)})^2}+\frac{1}{\sigma^{(M)}} & \frac{-(Q^0-M^0) (Q^1-M^1)}{R^2 (\sigma^{(M)})^2} & \frac{-(Q^0-M^0) Q^2}{R^2 (\sigma^{(M)})^2} & \frac{-(Q^0-M^0) Q^3}{R^2 (\sigma^{(M)})^2}\\
 \frac{-(Q^0-M^0) (Q^1-M^1)}{R^2 (\sigma^{(M)})^2} & \frac{(Q^1-M^1)^2}{R^2 (\sigma^{(M)})^2}-\frac{1}{\sigma^{(M)}} & \frac{(Q^1-M^1) Q^2}{R^2 (\sigma^{(M)})^2} & \frac{(Q^1-M^1) Q^3}{R^2 (\sigma^{(M)})^2} \\
 \frac{-(Q^0-M^0) Q^2}{R^2 (\sigma^{(M)})^2} & \frac{(Q^1-M^1) Q^2}{R^2 (\sigma^{(M)})^2} & \frac{(Q^2)^2}{R^2 (\sigma^{(M)})^2}-\frac{1}{\sigma^{(M)}} & \frac{Q^3 Q^2}{R^2 (\sigma^{(M)})^2} \\
 \frac{-(Q^0-M^0) Q^3}{R^2 (\sigma^{(M)})^2} & \frac{(Q^1-M^1) Q^3}{R^2 (\sigma^{(M)})^2} & \frac{Q^3 Q^2}{R^2 (\sigma^{(M)})^2} & \frac{(Q^3)^2}{R^2 (\sigma^{(M)})^2}-\frac{1}{\sigma^{(M)}}
 \end{array}
\right)\\
\la{6-1}
\eea
where
\bea\nn
\sigma^{(M)}&=&1-{(Q^0(z)-M^0)^2-(Q^1(z)-M^1)^2-(Q^2(z))^2-(Q^3(z))^2\over R^2}\\
\la{6-2}&=&1-{1\over R^2}[(Q^0(z)-M^0)^2-(Q(z)\cos \Theta-M^1)^2-Q(z)^2\sin^2\Theta],
\eea
where $Q^1(z)=Q(z)\cos\Theta$ and $(Q^2(z))^2+(Q^3(z))^2=Q(z)^2-Q^1(z)^2=Q(z)^2\sin^2\Theta$ have been used. From Eqs.(\ref{6-1}) (\ref{6-2}), we have
\bea\la{6-3}
\det B^{(M)}_{\mu\nu}(\mathcal{Q})=-(\sigma^{(M)})^{-5}.
\eea
From Eq.(\ref{A-56}), we have
\bea\la{6-4}
\lambda_1=-\sigma^{(M)}\left(1+{1\over R^2}[(Q(z)\cos\Theta-M^1)^2+Q(z)^2\sin^2\Theta]\right),~~\lambda_2=\lambda_3=-\sigma^{(M)}.
\eea
We now focus on the derivations of $\alpha$ in this case. In section 5, we presented the procedure for calculating $\alpha$ step by step in detail based on $ B_{\mu\nu}^{(M)}(\mathcal{Q})|_{(Q^2=Q^3=0)}$. Though the full $ B_{\mu\nu}^{(M)}(\mathcal{Q})$ (\ref{6-1}) is more complex than $ B_{\mu\nu}^{(M)}(\mathcal{Q})|_{(Q^2=Q^3=0)}$ (\ref{AB1}), the calculations in section 5 can be repeated smoothly. The resulting $\alpha_z(\Omega)$-expression (\ref{104}) keeps invariant except the $\{B_{\mu\nu}^{(M)}(\mathcal{Q}),\;\sigma^{(M)},\;\lambda_i\}|_{(Q^2=Q^3=0)}$ in the formula should be replaced by $\{B_{\mu\nu}^{(M)}(\mathcal{Q}),\;\sigma^{(M)},\;\lambda_i\}|_{(Q^i\neq0)}$ with $(i=1,2,3)$.
Namely we have
\bea\la{6-5}
\alpha_z(\Omega) =\alpha {{1\over\sqrt{\sigma^{(M)}}}+{(Q^0-M^0)^2\over R^2\sigma^{(M)}}\left({1\over 1+\sqrt{\sigma^{(M)}}}\right)\over
\sqrt{-\det(B_{\mu\nu}^{(M)}(\mathcal{Q}))} \sqrt{-\lambda_1\lambda_2\lambda_3}},~~{\rm with}~~ \alpha={e^2\over \hbar c},
\eea
where $\sigma^{(M)},\;\det B_{\mu\nu}^{(M)}(\mathcal{Q}),\;\lambda_i$ are given in Eqs.(\ref{6-2}), (\ref{6-3}) and (\ref{6-4}) respectively. The corresponding $\alpha$-varying formula reads
\bea\la{6-6}
{\Delta\alpha\over\alpha_0}=  {{1\over\sqrt{\sigma^{(M)}}}+{(Q^0-M^0)^2\over R^2\sigma^{(M)}}\left({1\over 1+\sqrt{\sigma^{(M)}}}\right)\over
N_0\sqrt{-\det(B_{\mu\nu}^{(M)}(\mathcal{Q}))} \sqrt{-\lambda_1\lambda_2\lambda_3}}-1,
\eea
where $N_0$ is given in (\ref{109}).

\begin{figure}[h]
\begin{center}
\includegraphics[scale=0.8]{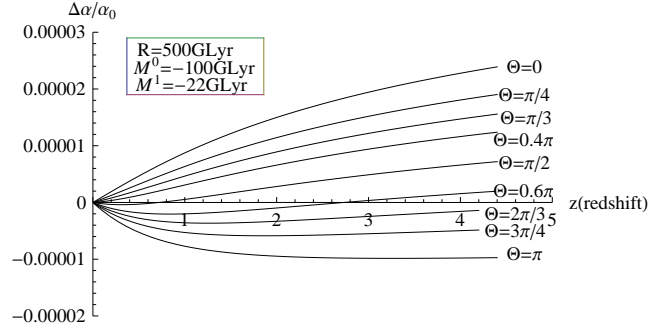}
\caption{\label{fig8} \small $\alpha$-varying $\Delta\alpha/\alpha_0(z,\Theta)$ in the region of $0<z<4.5$ and $0\leq \Theta \leq \pi$. The parameters $\{R,\;M^0,\;M^1\}$ are shown in the figure. }
\end{center}
\end{figure}

The $\alpha$-varyings $\Delta\alpha/\alpha_0 (z,\Theta)$ are shown in Fig.\ref{fig8} by using (\ref{6-6}) in which the curves correspond to $z$ from 0 to 4.5 and $\Theta=\{0,\;\pi/4,\;\pi/3,\;0.4\pi,\;\pi/2,\;0.6\pi \;2\pi/3,\;3\pi/4,\;\pi\}$ respectively.  We can see from the figure that: (i) When $z$ were fixed, $\Delta\alpha/\alpha_0$ decreases along with $\Theta$ increases from 0 to $\pi$; (ii) In regions of $\{0\leq\Theta<0.4\pi\}$ and $\{0.6\pi<\Theta\leq\pi\}$  ,$\alpha$ vary special spatially.
That is, $\alpha$ could be smaller in one direction in the sky yet larger in the opposite direction
at the time of absorption. This feature is consistent to the observations in Keck and VLT reported by \cite{Webb1} and \cite{Webb0};  (iii) When $\Theta\sim \{0.5\pi,\;0.6\pi,\;0.7\pi\}$, the observation results of $\alpha$-variations $\Delta \alpha/\alpha_0$ are nearly null.

\begin{figure}[h]
\begin{center}
\includegraphics[scale=0.6]{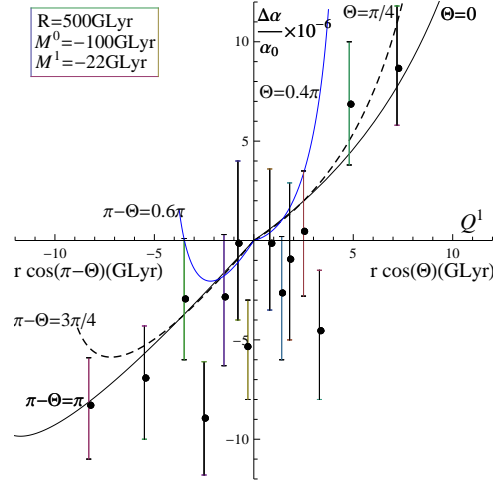}
\caption{\label{fig10} \footnotesize Curves of $\Delta\alpha/\alpha_0(z,\Theta)$ vs $A\;r\cos\Theta$. Two  solidline curves and one dotted curve are shown. One of the solidline curves corresponds to $\Theta=0$ and the other is for $\Theta=0.4\pi$. The dotted line curve corresponds to $\Theta=\pi/4$. The horizontal axis shows projection of atom's ``distance" $r(z)\equiv c t(z)$ onto $Q^1$ axis. $t(z)$ and cosmology parameters $(H_0,\;\Omega_M,\;\Omega_\Lambda)$ were given in (\ref{newLa1}) and (\ref{newLa4}) respectively. The data reported by \cite{Webb1} are plotted with error bars. The parameters $\{R,\;M^0,\;M^1\}$ are shown in the figure. }
\end{center}
\end{figure}

In order to show $\Delta\alpha/\alpha_0$'s dipolar behavior more explicitly, we further plot figure of $\Delta\alpha/\alpha_0$ vs $A\;r\cos\Theta$ in Fig.\ref{fig10}. Three theoretical prediction curves corresponding to $\Theta=\{0,\;\pi/4,\;0.4\pi\}$ and the experiment observation data reported by \cite{Webb1} are shown in the figure for comparison. It can been seen that the three curves are approximately close to each other in the region of $r\cos\Theta=\{-2.5{\rm GLyr}, \;2.5{\rm GLyr}\}$, and their average gradient is about $A\simeq 1.0\times 10^{-6}$GLyr$^{-1}$. This theoretical prediction value is consistent with the observation's $(1.1\pm 0.25)\times 10^{-6}{\rm GLyr}$ reported in \cite{Webb1}.
However, for absorbing systems with $\Theta\simeq 0.4\pi$ and $|r\cos\Theta|\geq 3.5{\rm GLyr}$, the curve with $\Theta=0.4\pi$ in Fig.\ref{fig10} indicates that $\Delta\alpha/\alpha_0(\Theta\sim 0.4\pi)\neq -\Delta\alpha/\alpha_0 (\Theta\sim (\pi-0.4\pi))$. This means that the dipole term (i.e., the term $\propto\cos\Theta$) in the expansion of $\Delta\alpha/\alpha_0$ is no longer dominating. For the absorbing systems with $0.4\pi<\Theta< 0.5\pi$ the situations is also similar. Observational experiments to check such predictions is called for.

\section{Summary and Discussions}
The spacetime variations of fine-structure constant $\alpha\equiv e^2/\hbar c$ in cosmology is a new phenomenon beyond the standard model of physics. To reveal the meaning of such new physics is of utmost importance to a complete understanding of fundamental physics. The main conclusion of this paper is that the phenomenon of $\alpha$-varying cosmologically with dipole mode dominating is due to the de Sitter (or anti de Sitter) spacetime symmetry in an extended special relativity called de Sitter invariant spacial relativity (dS-SR). Specifically, the logic that leads to this conclusion are summarized as  follows:
\begin{enumerate}
\item The Keck+VLT data that imply varying $\alpha$ are results of measuring spectra of atoms (or ions) in distant absorption clouds. So it is legitimate to use the electron wave equation of atoms (typically, the Hydrogen atom) to discuss this issue.
\item  As usual,  QM equations for spectra in atoms are defined in inertial coordinate systems to avoiding ambiguities caused by inertial forces. So, it is necessary to take the local inertial coordinate systems in FRW Universe for discussing both laboratory atoms and the distant.
\item When Einstein's cosmology constant $\Lambda\equiv 3/R^2\neq 0$, the metric in the local inertial coordinate systems in FRW Universe has to be Beltrami metric $B_{\mu\nu}(x)$ or M-Beltrami metric $B^{(M)}_{\mu\nu}(x)$, but cannot be Minkowski metric $\eta_{\mu\nu}$.
\item Since there exist both temporal and spatial variations for $\alpha$ in cosmology, $B^{(M)}_{\mu\nu}(x)$ is suitable. The de Sitter pseudo-radius parameter $R$ and the Minkowski point parameters $\{M^0,\;M^1\}$ are expected to be determined by fitting to the observations.
\item As usual, dS-Dirac equation for hydrogen can always be reduced to spectrum equation of hydrogen. In this way \cite{Yan0,Yan01} both coefficient of Dirac-kinetic energy operator term $-i\overrightarrow{\alpha}\cdot\nabla$ (i.e., $\hbar_z(\Omega)c$) and coefficient of Coulomb potential term $-1/r$ (i.e., $e^2_z(\Omega)$) are derived explicitly. Then $\alpha_z(\Omega)=e^2_z(\Omega)/\hbar_z(\Omega)c$ and $\Delta\alpha/\alpha_0=(\alpha_z(\Omega)-\alpha_0)/\alpha_0$ can be calculated.
\item According to \cite{Webb1,Webb0}, we focuss the best-fit direction about right ascension $\sim 17.5\;h$, declination $\sim -61 \deg$, and calculate $\Delta\alpha/\alpha_0$ in this region. Comparing the theoretical prediction with the observation results reported in \cite{Webb1,Webb0}, the model's parameters are determined: $R\simeq 500{\rm GLyr},\;M^0\simeq -100{\rm GLyr}$ and $M^1\simeq -22{\rm GLyr}$. Surprisingly but not strangely, the amazing observational discover about $\Delta\alpha/\alpha_0$-dipole in \cite{Webb1,Webb0} is reproduced by this dS-SR theoretical model. This is a main result of this paper, which could be thought as a nontrivial evidence to support dS-SR.
\item The $\alpha$-varying in the whole sky have also been studied in this model with the same parameters. The results are generally in agreement with the estimations in \cite{Webb1,Webb0}.
\end{enumerate}
To get things more straight, let's go back temporarily to the beginning of this paper again. If the theory of SR is exactly Einstein's SR (E-SR) with metric $\eta_{\mu\nu}$, the $\alpha$ will not vary over cosmic time, i.e., $\Delta\alpha/\alpha =0$. However, the observations among cosmological distances go directly against it, i.e., $(\Delta\alpha/\alpha)_{obs}\neq 0$. Even though people would be open-minded to face this big challenge \cite{Uzan},  the simplest answer to the puzzle seems to be that E-SR might not be exact for cosmological spacetime scale, or E-SR needs to be extended. To the best of authors' knowledge, the most natural consideration to extend E-SR in the framework of SR is the works due to Dirac(1935)-In\"{o}n\"{u} and Wigner(1953)-G\"{u}rsey and T.D. Lee (1968)-Lu, Zuo and Guo (1974) \cite{Dirac35,Wigner,Lee,look,Lu74}. Those works produced the theory of de Sitter invariant special relativity (dS-SR). Using the dS-SR in this  paper, we correctly work out the prediction of $\Delta\alpha/\alpha_0$ which is consistent with the data reported by \cite{Webb1,Webb0}. Hence the puzzle of $\alpha$-varying over cosmological time could be considered solved. Our approach could be considered a very simple answer  to the problem, if not the simplest answer.

The relativity principle problem in curved spacetime with constant curvature has been solved in dS-SR in \cite{look,Lu74} via introducing Beltrami metric (see \cite{Ours} and the Appendix of \cite{Sun}), and hence E-SR is the limit of dS-SR with $|R|\rightarrow \infty$. Since $|R|$-value determined in the present paper is fortunately cosmology huge ($\sim 500$GLyr), we argue that such dS-SR would not contradict the experiments verified E-SR within the error bands. Furthermore, the many-multiplet (MM) method \cite{Webb99,Webb992} used in \cite{Webb1,Webb0} is itself $R^{-1}$-free. That the $R$ is so huge (or almost infinity) means that the estimates to $\Delta\alpha/\alpha$ in \cite{Webb1,Webb0} are reliable approximately.

\vskip0.3in
\noindent {\Large\bf Acknowledgement:}
Authors would like to acknowledge Yong-Shi Wu for stimulating discussions on this topic. We also thank Gui-Jun Ding, Wen Zhao and Zi-Jia Zhao for much help. This work is Supported in part by National Natural Science Foundation of China under Grant No. 11375169.

\vskip0.5in

\appendix
\section{Beltrami Metric and de Sitter Invariant Special Relativity}

In this Appendix we briefly interpret the Beltrami metric and the de Sitter invariant Special Relativity (dS-SR).

\begin{enumerate}
\item Beltrami metric:

We derive the expression of Beltrami metric (\ref{00}) in the text.
We consider a 4-dimensional pseudo-sphere (or hyperboloid) $\mathcal{S}_\Lambda$ embedded in a 5-dimensional Minkowski spacetime with metric $\eta_{AB} =diag(1,-1,-1,-1,-1)$:
\bea\nn
\mathcal{S}_\Lambda :&&\eta_{AB}\xi^A\xi^B=-R^2,\\
\la{3-1}&& ds^2=\eta_{AB}d\xi^Ad\xi^B,
\eea
where index $A,\;B=\{0,1,2,3,5\} $, $R^2:=3\Lambda^{-1}$ and $\Lambda$ is the cosmological constant.
$\mathcal{S}_\Lambda$ is also called de Sitter pseudo-spherical surface with radii $R$. Defining
\bea\la{3-2}
x^\mu:=R{\xi^\mu\over \xi^5},~~{\rm with}~~\xi^5\neq 0,~{\rm and}~\mu=\{0,1,2,3\}.
\eea
and treating $x^\mu$ are Cartesian-type coordinates of a 4-dimensional spacetime with metric $g_{\mu\nu}(x)\equiv B_{\mu\nu}(x)$, denoting this 4-dimensional spacetime as $\mathcal{B}_\Lambda$ (call it Beltrami spacetime), we derive $B_{\mu\nu}(x)$ by means of the geodesic projection of $\{\mathcal{S}_\Lambda\mapsto \mathcal{B}_\Lambda\}$ (see Figure \ref{Fig11}).
\begin{figure}[ht]
\begin{center}
\includegraphics[width=0.5\textwidth]{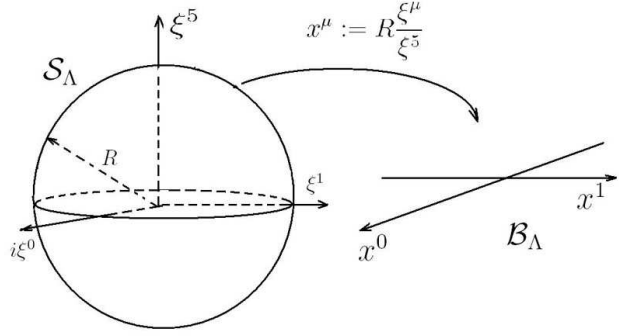}
\caption{\label{Fig11} \small Sketch of the geodesic projection from de Sitter pseudo-spherical surface $\mathcal{S}_\Lambda$ to the Beltrami spacetime $\mathcal{B}_\Lambda$ via Eq.(\ref{3-2}).}
\end{center}
\end{figure}
From the definition (\ref{3-1}), we have
\bea\nn
ds^2&=&\eta_{AB}d\xi^A d\xi^B|_{\xi^{A,B}\in \mathcal{S}_\Lambda}\\
\nn&=&\eta_{\mu\nu}d\xi^\mu d\xi^\nu-(d\xi^5)^2\\
\la{3-3}&:=&B_{\mu\nu}(x)dx^\mu dx^\nu .
\eea
Since $\xi^{A,B}\in \mathcal{S}_\Lambda $, and from (\ref{3-2}) and (\ref{3-1}), it is easy to obtain:
\bea\nn
&&\xi^\mu={x^\mu \over R}\xi^5,~~d\xi^\mu={1\over R}(\xi^5dx^\mu+x^\mu d\xi^5),~~ (\xi^5)^2={R^2\over \sigma(x)},\\
\nn && d\xi^5=\eta_{\mu\nu} {\xi^\mu\over \xi^5}d\xi^\nu={1\over R}\eta_{\mu\nu}x^\mu d\xi^\nu
={\eta_{\mu\nu}x^\mu dx^\nu\over \xi^5\sigma(x)^2},
\eea
where
\bea\la{333}
\sigma(x)= 1-{\eta_{\mu\nu}x^\mu x^\nu \over R^2}.
\eea
Substituting them into Eq.(\ref{3-3}), we have
\bea\nn
ds^2={\eta_{\mu\nu}dx^\mu dx^\nu\over \sigma(x)}+{(\eta_{\mu\nu}x^\mu dx^\nu)^2\over R^2 \sigma(x)^2}:= B_{\mu\nu}(x)dx^\mu dx^\nu.
\eea
Then, we obtain the Beltrami metric as follows
\bea\la{3-4}
B_{\mu\nu}(x)={\eta_{\mu\nu}\over \sigma(x)}+{\eta_{\mu\lambda}x^\lambda \eta_{\mu\rho}x^\rho\over R^2 \sigma(x)^2},
\eea
which is just Eq.(\ref{00}) in the text.
\item Inertial reference coordinates and principle of relativity: \\
The first Newtonian law is the foundation of the relativity. This law claims that the free particle moves with uniform velocity and along straight line.
There exist systems of reference in which the first Newtonian motion law holds. Such reference systems are defined to be {\it inertial}. And the Newtonian motion law is always called {\it the inertial moving law}. If two reference systems move uniformly relative to each other, and if one of them is an inertial system, then clearly the other is also inertial.
Experiment, e.g., the observations in the Galileo-boat which moves uniformly, shows that the so-called {\it principle of relativity} is valid. According to this principle all the law of nature are identical in all inertial systems of reference.

\noindent {\it Theorem 1}: The motion of particle with mass $m_0$ and described by the following Lagrangian
\begin{equation}\label{A1}
L_{Newton}={1\over2}m_0\mathbf{v}^2={1\over2}m_0\dot{\mathbf{x}}^2
\end{equation}
satisfy the first Newtonian motion law, or the motion is {\it inertial}. In (\ref{A1}), the Cartesian expression
of the velocity is as follows
\begin{equation}\label{A2}
\mathbf{v}\equiv \dot{\mathbf{x}},~~{\rm and}~~\mathbf{x}=x^1\mathbf{i}+x^2\mathbf{j}+x^3\mathbf{k},
\end{equation}
where $\mathbf{i}\cdot\mathbf{i}=\mathbf{j}\cdot\mathbf{j}=\mathbf{k}\cdot\mathbf{k}=1$, and $\mathbf{i}\cdot\mathbf{j}=\mathbf{i}\cdot\mathbf{k}=\mathbf{j}\cdot\mathbf{k}=0$.

{\it Proof}: By means of the Euler-Lagrangian equation
\begin{equation}\label{A3}
{\pa L \over \pa x^i}={d \over dt} {\pa L \over \pa \dot{x}^i},~~{\rm or}~~{\pa L \over \pa \mathbf{x}}={d \over dt} {\pa L \over \pa \dot{\mathbf{x}}}
\end{equation}
(where $\pa /\pa \mathbf{x}\equiv \nabla:=(\pa /\pa x^1)\mathbf{i}+(\pa /\pa x^2)\mathbf{j}+(\pa /\pa x^3)\mathbf{k}$ and etc) and $L=L_{Newton}$ we obtain
\begin{equation}\label{A4}
\ddot{x}^i=0,~~~~\dot{x}^i=v^i=constant,~~{\rm or}~~\dot{\mathbf{x}}=\mathbf{v}=constant.~~~~QED.
\end{equation}

\noindent {\it Theorem 2}: The motion of particle in Minkowski spacetime described by
\begin{equation}\label{A5}
L_{Einstein}=-m_0c{ds\over dt}=-m_0c{\sqrt{\eta_{\mu\nu}dx^\mu dx^\nu}\over dt}=-m_0c^2\sqrt{1-{\dot{\mathbf{x}}^2\over c^2}}
\end{equation}
is inertial.

The proof is the same as above, because both $L_{Newton}$ and $L_{Einstein}$ are coordinates
$x^i$-independent. Generally, any $\mathbf{x}$-free and time $t$-free Lagrangian functions $L(\dot{\mathbf{x}})$ can always reach the result of (\ref{A4}). However,
when Lagrangian function is time-dependent that rule will become invalid.
A useful example is as follows:
\begin{eqnarray}\label{A6}
L_{\Lambda}(t,\mathbf{x},\dot{\mathbf{x}})
=-m_0c^2 \sqrt{3/\Lambda} \sqrt{3/\Lambda (c^2- \dot{\mathbf{x}}^2)-\mathbf{x}^2\dot{\mathbf{x}}^2+(\mathbf{x}\cdot\dot{\mathbf{x}})^2+c^2(\mathbf{x}-\dot{\mathbf{x}}t)^2  \over c^2 (3/\Lambda +\mathbf{x}^2-c^2t^2)^2},
\end{eqnarray}
where a constant $\Lambda\neq 0$. The stick-to-itive readers can verify the following identity via straightforward calculations from (\ref{A6}):
\bea\la{A7}
{\pa L_{\Lambda} \over \pa \mathbf{x}}={\pa \over \pa t} {\pa L_{\Lambda} \over \pa \dot{\mathbf{x}}}+\left(\dot{\mathbf{x}}\cdot{\pa \over \pa\mathbf{x}}\right) {\pa L_{\Lambda} \over \pa \dot{\mathbf{x}}}.
\eea
Noting that the Euler-Lagrange equation (\ref{A3}) reads
\bea\la{A8}
{\pa L_{\Lambda} \over \pa \mathbf{x}}={d \over dt} {\pa L_{\Lambda} \over \pa \dot{\mathbf{x}}} ={\pa \over \pa t} {\pa L_{\Lambda} \over \pa \dot{\mathbf{x}}}+\left(\dot{\mathbf{x}}\cdot{\pa \over \pa\mathbf{x}}\right) {\pa L_{\Lambda} \over \pa \dot{\mathbf{x}}}+ \left(\ddot{\mathbf{x}}\cdot{\pa \over \pa\dot{\mathbf{x}}}\right) {\pa L_{\Lambda} \over \pa \dot{\mathbf{x}}},
\eea
and substituting (\ref{A7}) to (\ref{A8}), we have
\bea\la{A9}
\left(\ddot{\mathbf{x}}\cdot{\pa \over \pa\dot{\mathbf{x}}}\right) {\pa L_{\Lambda} \over \pa \dot{\mathbf{x}}}=0.
\eea
Since
\bea\la{A10}
\|{\pa \over \pa\dot{\mathbf{x}}} {\pa L_{\Lambda} \over \pa \dot{\mathbf{x}}}\|\equiv
\det \left({\pa^2L_\Lambda\over \pa x^i\pa x^j}\right)\neq 0
\eea
we have
\begin{equation}\label{A11}
\ddot{\mathbf{x}}=0,~~~~\dot{\mathbf{x}}=\mathbf{v}=constant,
\end{equation}
which indicates that the particle motion described by Lagrangian function (\ref{A6}) is inertial, and the first Newton motion law holds. Thus, the corresponding inertial reference systems can be built. Noting
\bea\la{A12}
\lim_{\Lambda\rightarrow 0}L_{\Lambda}=L_{Einstein},
\eea
it is essential and remarkable that a new kind of Special Relativity based on $L_\Lambda$ (\ref{A6}) serving as an extension of the Einstein's Special Relativity (E-SR) may exist.

\item de Sitter invariant Special Relativity (dS-SR):\\
Following the Landau-Lifshitz formulation of Lagrangian \cite{Landau} (see (\ref{A5})), we examine the motion of free particle in the spacetime with Beltrami metric (\ref{3-4}). From Eq.(\ref{4}) in text
\begin{equation}\label{B1}
 L_{dS}=-m_0c \frac{ds}{dt}
 =-m_0c{\sqrt{B_{\mu\nu}(x)dx^\mu dx^\nu}\over dt}=-m_0c{\sqrt{B_{\mu\nu}(x)\dot{x}^\mu \dot{x}^\nu}},
 \end{equation}
we derive its expression in Cartesian coordinates.
Setting up the time
$t=x^0/c$,   $B_{\mu\nu}(x)$ can be rewritten as follows
\begin{eqnarray}\nn
ds^2\hskip-0.06in &=&\hskip-0.06in  B_{\mu\nu}(x) dx^\mu dx^\nu
=\widetilde{g}_{00}d(ct)^2+\widetilde{g}_{ij}\left[(dx^i+N^id(ct))
(dx^j+N^jd(ct))\right]\\ \label{3-19app} &=& c^2 (dt)^2
\left[\widetilde{g}_{00} +\widetilde{g}_{ij}({1\over
c}\dot{x}^i+N^i) ({1\over c}\dot{x}^j+  N^j)\right],
\end{eqnarray}
where
\begin{eqnarray}\label{3-20}
\widetilde{g}_{00}&=&{R^2\over \sigma(x) (R^2-c^2t^2)},\\
\label{3-21} \widetilde{g}_{ij}&=&{\eta_{ij}\over \sigma (x)}+
{1\over
R^2\sigma(x)^2}\eta_{il}\eta_{jm}x^lx^m,\\
\label{3-22} N^i&=&{ctx^i \over R^2-c^2t^2}.
\end{eqnarray}
Substituting eqs.(\ref{3-19app})--(\ref{3-22}) into (\ref{B1}), we
obtain the Lagrangian for free particle in $ \mathcal{B}_{\Lambda} $:
\begin{equation}\label{3-23}
 L_{dS}=-m_0c^2 \sqrt{\widetilde{g}_{00} +\widetilde{g}_{ij}({1\over
c}\dot{x}^i+N^i) ({1\over c}\dot{x}^j+  N^j)}.
 \end{equation}
By using Cartesian notations (\ref{A2}) and expressions of (\ref{333}) (\ref{3-20}) (\ref{3-21}) (\ref{3-22}), the explicit expression of Lagrangian (\ref{3-23}) is:
\bea\nn
L_{dS}&\hskip-0.1in=&\hskip-0.1in -m_0c^2\left[{R^4\over (R^2+\mathbf{x}^2-c^2t^2)(R^2-c^2t^2)}\right.\\
\nn && +\left.{-R^2\over R^2+\mathbf{x}^2-c^2t^2}\right.\left({\dot{\mathbf{x}}^2\over c^2}+{c^2t^2\mathbf{x}^2\over (R^2-c^2t^2)^2}+{2t(\mathbf{x}\cdot\dot{\mathbf{x}})\over R^2-c^2t^2}\right)\\
\nn && +\left.{R^2\over (R^2+\mathbf{x}^2-c^2t^2)^2}\left({\dot{\mathbf{x}}\cdot\mathbf{x}\over c}+{ct\mathbf{x}^2\over R^2-c^2t^2}\right)^2\right]^{1/2}\\
\la{3-26} &=& -m_0c^2 R \sqrt{R^2 (c^2- \dot{\mathbf{x}}^2)-\mathbf{x}^2\dot{\mathbf{x}}^2+(\mathbf{x}\cdot\dot{\mathbf{x}})^2+c^2(\mathbf{x}-\dot{\mathbf{x}}t)^2  \over c^2 (R^2+\mathbf{x}^2-c^2t^2)^2},
\eea
where $\mathbf{x}^2=(\mathbf{x}\cdot\mathbf{x})$. Noting $R^2=3/\Lambda$ (see, e.g., Eq.(15) in Ref. \cite{Sun}), and comparing $L_{dS}$ with $L_{\Lambda}(t,\mathbf{x},\dot{\mathbf{x}})$ of (\ref{A6}), we find
\bea\la{B10}
L_{dS}=L_{\Lambda}(t,\mathbf{x},\dot{\mathbf{x}})=-m_0c{\sqrt{B_{\mu\nu}(x)\dot{x}^\mu \dot{x}^\nu}},
\eea
which is the Lagrangian for free particle mechanics of dS-SR. Since (\ref{A12}), when $|R|\rightarrow \infty$, the dS-SR goes back to E-SR.

\item de Sitter transformation to preserve Beltrami metric $B_{\mu\nu}$:\\
In \cite{Ours} (see Eqs. (41)--(43) in \cite{Ours}), we have shown that under Lu-Zou-Guo (LZG) transformation (see also Eq.(\ref{a-7}) below)
preserves Beltrami metric $B_{\mu\nu}$. When  space rotations were neglected temporarily  for simplify, the
transformation both due to a Lorentz-like boost and a
space-transition in the $x^1$ direction with  parameters
$\beta=\dot{x}^1/c$ and $a^1$ respectively and due to a time
transition with  parameter $a^0$ can be explicitly written as
follows:
\begin{eqnarray}\label{general transformation}
\begin{array}{rcl}
t\rightarrow \tilde{t}&=& \frac{\sqrt{\sigma(a)}}{c \sigma(a,x)}
\gamma \left[ct-\beta x^1-a^0+ \beta a^1 +\frac{a^0-\beta
a^1}{R^2}\frac{a^0 ct-a^1 x^1-(a^0)^2 +(a^1)^2 }
{ \sigma(a)+\sqrt{\sigma(a)}} \right] \\
 x^1\rightarrow \tilde{x}^1&=& \frac{\sqrt{\sigma(a)}}{
\sigma(a,x)}\gamma \left[ x^1-\beta ct +\beta a^0 -a^1 +\frac{a^1-
\beta a^0}{R^2}
\frac{a^0 ct-a^1 x^1-(a^0)^2 +(a^1)^2}{ \sigma(a)+\sqrt{\sigma(a)}}\right]\\
 x^2\rightarrow
\tilde{x}^2&=&\frac{\sqrt{\sigma(a)}}{\sigma(a,x)}x^2 \\
 x^3\rightarrow
\tilde{x}^3&=&\frac{\sqrt{\sigma(a)}}{\sigma(a,x)}x^3
\end{array}
\end{eqnarray}
where $\gamma=1/\sqrt{1-\beta^2}$. It is easy to check when $R\rightarrow \infty$ the above
transformation goes back to Poincar\'{e} transformation (or
inhomogeneous Lorentz group $ISO(1,3)$ transformation) in E-SR.

\item Conserved Noether charges of $SO(4,1)$ of dS-SR:\\
The external spacetime symmetry of dS-SR is $SO(4,1)$. According to Neother theorem, the corresponding 10-Noether charges are energy $E$, momentums $p^i$, boost charges $K^i$ and angular-momentums $L^i$. All have been derived in \cite{Ours}. The results are as follows
\begin{eqnarray}\label{503a}
\begin{array}{rcl}
  &&{\rm{ Noether}\;charges\;for\;Lorentz\;boost:\;} ~~
 K^i=m_0 \Gamma c (x^i- t \dot{x}^i) \\
 &&{\rm
Charges\;for\;space-transitions\;(momenta):}~~~  p^i=m_0 \Gamma \dot{x}^i, \\
 &&{\rm Charge\;for\;time-transition\;(energy): }~~~
 E= m_0 c^2 \Gamma \\
&&{\rm Charges\;for\;rotations\;in\;space\;(angular momenta):}~~~
L^i = \epsilon^{i}_{jk}x^{j}p^{k},
\end{array}
\end{eqnarray}
where the Lorentz factor of dS-SR is:
\begin{eqnarray} \label{new parameter}
 \Gamma ={1\over \sqrt{1-{\dot{\mathbf{x}}^2\over c^2}+{(\mathbf{x}\cdot \dot{\mathbf{x}})^2 -\mathbf{x}^2\dot{\mathbf{x}}^2\over c^2R^2}+{(\mathbf{x}-\dot{\mathbf{x}}t)^2\over R^2}}}.
\end{eqnarray}
It can be checked that $\dot{E}=\dot{p^i}=\dot{K^i}=\dot{L^i}=0$ under the equation of motion $\ddot{x}^i=0$ (or $\ddot{\mathbf{x}}=0$) \cite{Ours}.

\end{enumerate}



\section{Modified Beltrami Metric and de Sitter Invariant Special Relativity}

We provide a brief introduction to Modified Beltrami metric (M-Beltrami metric) and the corresponding dS-SR.

\begin{enumerate}
\item M-Beltrami metric: Eqs. (\ref{1-1}) and (\ref{1-2}) are the definition of M-Beltrami metric $B^{(M)}_{\mu\nu}(x)$. Being different from $B_{\mu\nu}(x)$, the coordinate components of Minkowski point for $B^{(M)}_{\mu\nu}(x)$ is $M^\mu$ instead of the origin of spacetime system $x^\mu$. Introducing notation
    \bea\la{a-1}
    y^\mu\equiv x^\mu-M^\mu
    \eea
then
\bea\la{a-2}
B^{(M)}_{\mu\nu}(x)=B_{\mu\nu}(y).
\eea
The Landau-Lifshitz action is
\bea\la{a-3}
S= -mc\int \sqrt{B^{(M)}_{\mu\nu}(x)dx^\mu dx^\nu}= -mc\int \sqrt{B_{\mu\nu}(y)dy^\mu dy^\nu},
\eea
where $ d M^\mu=0$ were used duo to constancy of $M^\mu$. The Lagrangian $L_{M-dS}$ reads
\bea\la{aa-3}
L_{M-dS}=-mc\sqrt{B^{(M)}_{\mu\nu}(x)\dot{x}^\mu\dot{x}^\nu}.
\eea
Then, from $\delta S=0$ and Eq.(\ref{A11}), we have
\bea\la{a-5}
{d^2{\mathbf y}\over d(y^0)^2}=0,
\eea
where $\mathbf{y}=\mathbf{x}-\mathbf{M},\;y^0=ct_y=ct-M^0$ (see Eq.(\ref{a-1})). Eq.(\ref{a-5}) becomes
\bea\la{a-6}
{d^2{\mathbf x}\over dt^2}\equiv \ddot{\mathbf x}=0,
\eea
which means that the free particle moves with uniform velocity and along straight line in the dS-SR based M-Beltrami metric. Consequently,  the first Newtonian law holds for $L_{M-dS}$ Eq.(\ref{aa-3}) and inertial coordinate frames are well defined.

\item Spacetime symmetries of M-Beltrami metric and the motion integrals. In \cite{Ours} (see Eqs. (41)--(43) in \cite{Ours}), we have shown that under Lu-Zou-Guo (LZG) transformation
\begin{eqnarray}\label{a-7}
y^{\mu} \;-\hskip-0.10in\longrightarrow\hskip-0.4in^{LZG}
 ~~ \tilde{y}^{\mu} &=& \pm \sigma(a)^{1/2} \sigma(a,x)^{-1}
(y^{\nu}-a^{\nu})D_{\nu}^{\mu}, \\
    \nonumber D_{\nu}^{\mu} &=& L_{\nu}^{\mu}+R^{-2} \eta_{\nu
\rho}a^{\rho} a^{\lambda} (\sigma
(a) +\sigma^{1/2}(a))^{-1} L_{\lambda}^{\mu} ,\\
\nonumber L : &=& (L_{\nu}^{\mu})\in SO(1,3), \\
\nonumber \sigma(y)&=& 1-{1 \over R^2}{\eta_{\mu \nu}y^{\mu} y^{\nu}}, \\
\nonumber \sigma(a,y)&=& 1-{1 \over R^2}{\eta_{\mu \nu}a^{\mu}
y^{\nu}},
\end{eqnarray}
the Beltrami metric transformation reads:
\begin{equation} \label{a-8}
 B_{\mu\nu}(y)\;-\hskip-0.10in\longrightarrow\hskip-0.4in^{LZG}
 ~~ ~\widetilde{B}_{\mu\nu}(\widetilde{y})={\pa y^\lambda \over \pa
 \widetilde{y}^\mu}{\pa y^\rho \over \pa
 \widetilde{y}^\nu}B_{\lambda\rho}(y)=B_{\mu\nu}(\widetilde{y}).
\end{equation}
(\ref{a-8}) leads to the invariance of action of (\ref{a-3}):
\begin{equation} \label{a-9}
 S \;-\hskip-0.05in-\hskip-0.05in\longrightarrow\hskip-0.4in^{LZG}
 ~ ~~\widetilde{S}=S.
\end{equation}
The corresponding Noether chargers or conserved motion integrals are as follows:
\begin{eqnarray}\label{a-10}
\begin{array}{rcl}
  &&{\rm{ Noether}\;charges\;for\;Lorentz\;boost:\;} ~~
 K^i=m \Gamma c (y^i- t_y {dy^i\over dt_y}) \\
 &&{\rm
Charges\;for\;space-transitions\;(momenta):}~~~  p^i=m \Gamma {dy^i\over dt_y}, \\
 &&{\rm Charge\;for\;time-transition\;(energy): }~~~
 E= m c^2 \Gamma \\
&&{\rm Charges\;for\;rotations\;in\;space\;(angular momenta):}~~~
L^i = \epsilon^{i}_{jk}y^{j}p^{k},
\end{array}
\end{eqnarray}
where the Lorentz factor of dS-SR is:
\begin{eqnarray} \label{a-11}
 \Gamma ={1\over \sqrt{1-{1\over c^2}\left({d\mathbf{y}\over dt_y}\right)^2+{1\over c^2R^2}\left[\left(\mathbf{y}\cdot {d\mathbf{y}\over dt_y}\right)^2 -\mathbf{y}^2\left({d\mathbf{y}\over dt_y}\right)^2\right]+{1\over R^2}\left(\mathbf{y}- t_y {d\mathbf{y}\over dt_y}\right)^2}}.
\end{eqnarray}
Using Eq.(\ref{a-1}), we have the expressions in $x$ frame:
\bea\la{a-12}
&&K^i=m \Gamma c [x^i-M^i- (t-M^0/c)\dot{x}^i], \\
\la{a-12}&& p^i=m \Gamma \dot{x}^i,\\
\la{a-13}&& E= m c^2 \Gamma,\\
\la{a-14}&& L^i = \epsilon^{i}_{jk}(x-M)^{j}p^{k},
\eea
and
\begin{eqnarray} \label{a-15}
 \Gamma ={1\over \sqrt{1-{\dot{\mathbf{x}}^2\over c^2}+{[(\mathbf{x}-\mathbf{M})\cdot \dot{\mathbf{x}}]^2 -(\mathbf{x}-\mathbf{M})^2\dot{\mathbf{x}}^2\over c^2R^2}+{[\mathbf{x}-\mathbf{M}-\dot{\mathbf{x}}(t-M^0/c)]^2\over R^2}}}.
\end{eqnarray}
It is straightforward to check that $\dot{E}=\dot{p^i}=\dot{K^i}=\dot{L^i}=0$ under the equation of motion $\ddot{x}^i=0$ (or $\ddot{\mathbf{x}}=0$) and $M^\mu=\text{const.}$.

\end{enumerate}



\newpage
\begin{thebibliography}{99}
\bibitem{dirac37}
P. A. M. Dirac, Nature {\bf139} (1937) 323.
\bibitem{Uzan} J.-P. Uzan, Rev. Mod. Phys. {\bf 75} 403 (2003).


\bibitem{Webb1} J. K. Webb, J. A. King, M. T. Murphy, V.V. Flambaum, R. F. Carswell, and M. B. Bainbridge, Phys. Rev. Lett., {\bf 107}, 191101 (2011).
\bibitem{Webb0}J. A. King, {\it et al.}, Mon.Not.Roy.Astron.Soc. {\bf 422} (2012) 3370-3413.
\bibitem{Webb99} J. K. Webb {\it et al.}, Phys. Rev. Lett. 82, 884
(1999).
\bibitem{Webb992} V. A. Dzuba, V.V. Flambaum, and J. K. Webb, Phys. Rev.
Lett. 82, 888 (1999).
\bibitem{Webb01} J. K. Webb {\it et al.}, Phys. Rev. Lett. 87, 091301
(2001).
\bibitem{Webb03} M. T. Murphy, J.K.Webb, and V.V. Flambaum, Mon. Not.
R. Astron. Soc. 345, 609 (2003).
\bibitem{Webb04}M. T. Murphy, V.V. Flambaum, J. K. Webb, V.V. Dzuba,
J. X. Prochaska, and A. M. Wolfe, Lect. Notes Phys. {\bf 648},
131 (2004).

\bibitem{Begen} J.D. Bekenstein, Phys. Rev. {\bf 25}, 1527 (1982); Phys.Rev. {\bf D66} 123514 (2002).

\bibitem{SBM} H.B. Sandvik ,  J.D. Barrow, and  J. Magueijo, Phys. Rev.
Lett.,  {\bf 88}, 031302 (2002).
\bibitem{Barrow} J.D. Barrow , Phys Rev.,  {\bf D71}: 083520 (2005);  J.D. Barrow ,  H.B.Sandvik, and J. Magueijo, Phys. Rev.
,  {\bf D65}: 063504 (2002),; {\it ibid}, 2002, {\bf D66}:
043515.
\bibitem{Olive} K.A. Olive, M.Peloso, and J.-P. Uzan, Phys. Rev., {\bf D 83}, 043509 (2011).

\bibitem{Dirac35} P.A.M. Dirac, Annal of Math. {\bf 36}, 657-689
(1935).

\bibitem{Wigner}E. In\"{o}n\"{u} and E. P. Wigner, Proc.Nat.Acad.Sci. {\bf 39}, 510-524 (1953);
H. Bacry and J.-M. L\'evy-Leblond, J. Math. Phys. {\bf 9}, 1605-1614 (1968).
\bibitem{Lee} F. G\"{u}rsey and T.D. Lee, Physics, {\bf 49}, 179 (1963).

\bibitem{look} K.H. Look (Q.K. Lu), {\it Why the Minkowski metric must be used?}, (1970), unpublished.
\bibitem{Lu74} K.H. Look, C.L. Tsou (Z.L. Zou) and H.Y. Kuo (H.Y. Guo), {\it Acta Physica Sinica}, {\bf 23} (1974) 225 (in Chinese).

\bibitem{Guo1} H.Y. Guo, C.G. Huang, Z. Xu, and B. Zhou, Phys. Lett. {\bf A 331} (2004) 1;  Mod. Phys. Lett. {\bf
A19} (2004) 1701; Chin. Phys. Lett. {\bf 22} (2005) 2477.

\bibitem{Ours} M.L. Yan, N.C. Xiao, W. Huang, S. Li,
Commun. Theor. Phys. {\bf 48} (2007) 27, arXiv:hep-th/0512319.


\bibitem{Yan0} M.L. Yan, Commum. Theor. Phys. {\bf 57} (2012) 930-952. arXiv:1004.3023 [physics.gen-ph].

\bibitem{Yan01} M.L. Yan, Commun. Theor. Phys. {\bf 62} (2014) 189-195.

\bibitem{Strange} P. Strange, ``{\it Relativistic Quantum Mechanics}", Cambrigdge University Press, (2008).

\bibitem{Sun} L.F. Sun, M.L.Yan, Y. Deng, W.Huang, S.Hu,  Modern Physics Letters {\bf A28}, (2013) 1350114. arXiv:1308.5222 [gr-qc].

\bibitem{Yan+} S.X. Chen, N.C. Xiao, M.L. Yan, Chinese Phys. {\bf C32}, 612 (2008).

\bibitem{Born} M. Born and V. Fock, Z. Phys., {\bf 51}, 165 (1928).
\bibitem{Messian} A. Messiah, {\it ``Quantum Mechanics I, II"}, North-Holland
Publishing Company, 1970.

\bibitem{Bayfield} J.E. Bayfield, {\it ``Quantum Evolution: An
Introduction to Time-Dependent Quantum Mechanics"}, John Wiley $\&$
Sons, Inc., New York, 1999.


\bibitem{Peebles} P.J.E. Peebles, Rev. of Mod. Phys. {\bf 75}, 559
(2009).

\bibitem{Pad}T.Padmanabhan, Phys. Rep. {\bf 380}, 235 (2003).
\bibitem{YanH} Mu-Lin Yan, Sen Hu, Wei Huang and Neng-Chao Xiao, {\it Mod.Phys.Lett.} {\bf A27}, 1250041, (2012). arXiv:1112.6217 [hep-ph].
\bibitem{L1} A.G. Riess, {\it et al.}, Astro. J. {\bf 116} 1009 (1998);  S. Perlmutter et al., Astrophys. J. {\bf 517}, 565 (1999) [astro-ph/9812133].
\bibitem{L2} N. Jarosik, et al.,  Astrophys. J. Suppl. {\bf 192}, 14 (2011); Planck Collaboration: P. A. R. Ade, {\it et al.}, {\it ``Planck 2013 results. XVI. Cosmological parameters"}, 	arXiv:1303.5076 [astro-ph.CO].

\bibitem{Weinberg} S.Weinberg, {\it ``Cosmology"}, Oxforrd University
Press Inc., New York, (2008).

\bibitem{Lambda} S. Weinberg, Rev. Mod. Phys. {\bf 61}, 1 (1989); T. Padmanabhan, Phys. Rep. {\bf 380}, 235 (2003).

\bibitem{Lambda1}  E. Komatsu, {\it et al.},
Astrophys.J.Suppl. {\bf 180} 330 (2009).

\bibitem{Hartle} see, e.g., J.B. Hartle, {\it ``Gravity, An Introduction to Einstein's General Relativity"}, Addison Wesley, (2003) {\it pp.119}.

\bibitem{Landau} L.D. Landau and E.M. Lifshitz, {\it The Classical
Theory of Fields}, (Translated from Russian by M. Hamermesh),
Pergamon Press, Oxford (1987).

\bibitem{Dent} T.Dent, S.Stern, and C.Wetterich, Phys. Rev. {\bf D78}, 103518 (2008).







\end{thebibliography}
\end{document}